\documentclass[fleqn,usenatbib]{mnras}

\usepackage{newtxtext,newtxmath,multirow,multicol,booktabs,siunitx}
\usepackage[T1]{fontenc}

\DeclareRobustCommand{\VAN}[3]{#2}
\let\VANthebibliography\thebibliography
\def\thebibliography{\DeclareRobustCommand{\VAN}[3]{##3}\VANthebibliography}

\usepackage{graphicx}	% Including figure files
\usepackage{amsmath}	% Advanced maths commands

\usepackage{capt-of}
\title[Semantic Segmentation of Stellar Streams]{From Trees to Transformers: Semantic Segmentation of Stellar Streams in Cosmological Simulations}

\author[V. Ekanayaka et al.]{
Viraj Ekanayaka,$^{1}$\thanks{E-mail: viraj.ekanayaka@sydney.edu.au}
Foivos Diakogiannis,$^{2}$
Smrithi Gireesh Babu$^{1}$
and Geraint F. Lewis$^{1}$
\\
$^{1}$Sydney Institute for Astronomy, School of Physics A28, The University of Sydney, NSW 2006, Australia\\
$^{2}$Technology, CSIRO, Kensington WA, Australia
}

\date{Accepted XXX. Received YYY; in original form ZZZ}

\pubyear{\the\year{}}

\begin{document}
\label{firstpage}
\pagerange{\pageref{firstpage}--\pageref{lastpage}}
\maketitle

\begin{abstract}
We present a benchmark test for a collection of Machine Learning (ML) approaches to classifying stellar streams in Milky Way-like cosmological simulation environments. We compare four frameworks, XGBoost enriched decision tree, voxel and pointcloud segmentation as well as a modified 6D-LitePT transformer. Importantly, we provide no temporal context and restrict available data to exclusively 6D phase space derived information. We show that all four models are able to isolate coherent stream-like structure in a synthetic testing data cloud. Along with visual inspection, we compare with a density based classification and find all four models to exceed its performance at recovering stream information. We then evaluate the original synthetic trained models on six independent Auriga level-3 cosmological simulations, in particular Au-24, using published system classifications as particle level reference labels. Across these halos, 6D-LitePT achieves the highest mean precision $0.617$ and Matthews Correlation Coefficient (MCC) of $0.326$, while Pointcloud achieves the highest mean recall $0.679$ with a MCC of $0.258$. These results demonstrate complementary strengths in stream recovery and contamination control, while identifying limitations in transfer from synthetic to cosmological populations. The framework provides a quantitative basis for developing stream selection methods for future observational applications.
\end{abstract}


%%%%%%%%%%%%%%%%%%%%%%%%%%%%%%%%%%%%%%%%%%%%%%%%%%

%%%%%%%%%%%%%%%%% BODY OF PAPER %%%%%%%%%%%%%%%%%%

\section{Introduction}

The current $\Lambda$CDM standard \citep{Blumenthal_1984} asserts a hierarchical structure formation model whereby larger galactic halos are primarily produced from smaller mergers \citep{Frenk_1988,Subramanian_2000}. As such, the tidal disruption of stellar structures such as dwarf galaxies and globular clusters are important events that provide insight into the history of galactic formation. Typically thin arcs of stars, stellar streams can be considered an intermediary phase of this process, wherein the progenitor remnants still retain phase space coherence before eventually mixing with the background density \citep{Cooper_2010,Helmi_2020}. These stellar streams encircle galactic halo systems and are highly sensitive to the potential they are disrupted in \citep{Ibata_1997,Odelkirchen_2001}. Thus, they also act as a valuable tool in probing the underlying dark matter distribution in these galactic systems \citep{Johnston_1999,Koposov_2010,Bovy_2016,Bonaca_2018}.

Motivated by what could be revealed, there has been an enthusiastic procession of studies delving into observational surveys in search of stellar streams. In particular SDSS \citep{SDSS}, DES \citep{DES} and Gaia \citep{GAIA} have proved to be rich troves of information \citep{Belokurov_2006,Shipp_2018,Bonaca_2025}. These have led to the discovery of several notable streams, including Sagittarius \citep{Ibata_1994,Ibata_1995} and GD-1 \citep{Grillmair_2006_gd1}. The more recent S5 Collaboration \citep{S5} have also continued to map out streams, endeavouring to obtain full 6D phase space and chemical abundance information for a suite of streams \citep{Li_2022}. Despite the success, finding stellar streams is a challenging task as they are Low Surface Brightness (LSB) objects that are difficult to identify and isolate amongst the noise of galactic systems. For this reason, we often turn to simulations as an alternative `sandbox' environment in which to better investigate the dynamics and morphology of these stellar stream structures \citep{Law_2010,Shipp_2021,Nibauer_2025}.

Whether it is observations or simulations, identifying and classifying stellar streams remains a non-trivial task. The most successful method thus far has been with simple visual classification. However, it is imperfect and perhaps more importantly, impractical at larger scales. Other methodologies have been put forward to solve the issue including density based approaches \citep{Ekanayaka_2025,Starkman_2025}, isolation in energy/angle approaches \citep{Hendel_2015} and other similar methodologies. With the advent of the Machine Learning (ML) era, we now have access to a more powerful tool that may be applied to the problem. There have been incursions in this regard such as \citet{BenitezWalz_2026}, but otherwise remains a particularly open field.

As a potential solution to the problem, we conduct a benchmark test using four separate ML pipelines with varied training and inference strategies. These include: 1. A gradient-boosted, enriched feature decision-tree via XGBoost, 2. Convolutional Neural Network (CNN) based Voxel segmentation method, 3. A PointNet++ \citep{Qi_2016,Qi_2017} inspired Pointcloud semantic segmentation method and 4. A hybrid LitePT \citep{Yue_2025} transformer model referred to as 6D-LitePT hereafter. We employ a training, validation and testing pipeline to quantify the effectiveness of each model at identifying stellar stream particles. Thereafter, we apply the models to an example independent dataset for a more accurate reflection of their practical capability. For this we chose the level-3 Auriga labelled suite \citep{Grand_2017,Grand_2024}, as high resolution Milky Way analogues.

This paper is structured as follows. In section \ref{sec:background} we provide a brief overview of stellar stream morphology and dynamics as well as some background context on the machine learning methodologies applied in this work. In section \ref{sec:benchmark_dataset} we present the benchmark dataset we use to train, validate and test our models. Section \ref{sec:ml_algorithms_benchmark_testing} details the four ML frameworks and presents the test results for the models. In section \ref{sec:independent_dataset_testing} we apply the models to the Auriga suite's Au-24 as well as Au-6,16,21,23,27 to test their effectiveness on independent datasets. We conclude in section \ref{sec:conclusion} and discuss future extensions to the presented work.

\section{Background}
\label{sec:background}
\subsection{Defining a `Stream'}
For the purpose of training an ML algorithm, we must tell it what a stellar stream is. However, this definition is not concrete. Typically, streams possess leading and trailing arms attached to the progenitor remnant core \citep{bullock_johnston,Amorisco_2015,Valenzuela_2024}. However, streams in more advanced stages of disruption lose distinction in this three-component structure and rather behave as a single homogeneous arc. Furthermore, these tails can be very diffuse, form multiple wraps, possess bifurcations or gaps \citep{Yavetz_2023,Nibauer_2024}. The morphological complexity of defining a `stream' often means additional dynamical perspective is required. 

Typically streams are dynamically cold, possessing velocity dispersions as low as $1-2$ km/s from globular cluster (GC) progenitors, while dwarf galaxy progenitors produce in the range of $5-20$ km/s \citep{Menker_2026}. Perhaps the most important dynamical marker of streams are their initial conditions for evolution. Streams are the result of slower tidal stripping passages in lower eccentric orbits. Vitally, an angular velocity component is required to ensure the core does not travel radially through the centre of the galactic potential. Such an orbit tends to completely disperse the core, leading instead to shell and plume-like structures. 
 
Observational data reveals additional layers of nuance introduced by interactions with other components in galactic systems. For example, \citet{Li_2021} discovered ATLAS and Aliqa Uma originate from a single progenitor, with the gap in the stream arising from the effects of the Large Magellanic Cloud and Sagittarius Dwarf. Similar perturbations have been discussed in a collection of other works such as \citep{Pearson_2015,Bonaca_2019,Cohen_2026}. However, these features are unpredictable and difficult to generalise. For the ML work presented here, we lean on the characteristic morphology and evolution history to define what we consider a stream. This is further discussed in section \ref{sec:benchmark_dataset}.

\subsection{ML Application Techniques}

The core concepts of ML have early origins with \citet{Mcculloch_1943} introducing the idea of an artificial neuron. Building on this, \citet{Rosenblatt_1958} proposed the theory of perceptrons, demonstrating the possibility of a neural network classifier. \citet{Rumelhart_1986} popularised back-propagation, allowing the development of multilayer neural networks.  However, without the computational resources necessary to effectively put it into action at that time, the development of the field progressed slowly. The 2010s saw the beginning of the revitalised ML era, with the demonstration of deep learning effectiveness via CNNs \citep{Krizhevsky_2012}, attention and transformers \citep{Vaswani_2017}.

Under the umbrella of ML, there exists several independent methodologies and approaches. One such algorithm is XGBoost \citep{Chen_2016}. While not truly deep learning, it is still able to `learn' the importance of particular data points to produce a classification. In essence, XGBoost takes a feature space and produces an ensemble of decision trees to distinguish between given classes. It excels at determining the ideal weights for any particular tree depending on how successful the tree is at separating data into the desired classes. Perhaps most importantly, XGBoost does not learn the feature space, and instead relies on user input of the metrics to produce a model with. The algorithm works by using gradient boosting \citep{Friedman_2001},
\begin{equation}
    F_m(x) = F_{m-1}(x) + \rho_m h(x; a_m),
    \label{eqn:gradient_boosting}
\end{equation}
to minimise the loss function:
\begin{equation}
    \mathcal{L}(\phi)=\sum_{i}l(\hat{y}_i, y_i)+\sum_{k}\Omega(f_k).
    \label{eqn:gradient_boosting_loss}
\end{equation}
Here, for input feature vector $x$, $F_{m-1}$ is the model after $m-1$ trees. The function $h(x; a_m)$ is the additional decision tree introduced at iteration $m$, with $a_m$ specifying its structure and leaf values. This is a correction value from the new tree’s output, which is added to the existing ensemble prediction to reduce the loss. $\rho_m$ is the weight for the correction. $\mathcal{L}(\phi)$ is the tree loss, while $l(\hat{y}_i, y_i)$ is the loss between $\hat{y}_i$ and target $y_i$ and $\sum_{k}\Omega(f_k)$ accounts for internal error in the modelling for the $k$th tree, $f_k$.

Moving to a more rigourous ML approach, we have semantic segmentation \citep{DeBrabandere_2017,Graham_2017,Foivos_2020,He_2021,Sarker_2024}. This range of techniques aim to separate classes of items from a given dataset. Importantly, unlike XGBoost, these techniques employ deep learning, producing learned representations that form the basis of their classification framework. For the work here, we specifically focus on Voxel and Pointcloud segmentation.

Voxel Segmentation \citep{Long_2015,Ronneberger_2015,Maturana_2015} involves the application of a CNN to a grid data structure. Although typically used in 2D image pixel processing, voxelisation may be utilised to extend the application to a 3D data cloud. The voxelisation process involves discretising an effectively continuous dataset into a boxed grid architecture. Each individual grid cube (voxel), inherits a summarised feature vector of the collection of particles contained within. Thus, the voxel becomes a blanket representation for the ensemble of particles inside its bounds. The downside here is that, in the case of a voxel containing particles from multiple classes, the dominating class will determine the class of the voxel. Once initialised, a 3D CNN may be run over the voxel grid box. Similar to 2D image learning, the CNN applies a filter across the box to extract pattern features between voxels. Sub-sampling with multiple iterations of box creation and voxelising provides a way to holistically learn from the entire data space.  

Although not strictly semantic segmentation, pointcloud segmentation effectively accomplishes the same general task of classifying separable classes in a given dataset. This is, however, approached in a fundamentally different manner. Established by PointNet and PointNet++ \citep{Qi_2016,Qi_2017}, pointcloud segmentation applies patch formation and feature aggregation to semantically classify individual particles. This process typically involves the allocation of a `centre' particle. A patch in Euclidean metric space is created around the selected `centre' particle. Multilayer Perceptron (MLP) layers are identically applied to each particle in the patch. The output from this is pooled into a summary patch used for further learning. PointNet++ takes this a step further, creating a hierarchical patch creation approach. 

The advent of attention and transformers has provided an additional avenue for investigation. Attention is defined as 
\begin{equation}
    \mathrm{Attention}(Q,K,V)=\mathrm{softmax}\left(\frac{QK^T}{\sqrt{d_k}}\right)V,
\end{equation}
where $\mathrm{softmax}$ provides weights and the matrices $Q$, $V$ and transpose matrix $K^T$ contain the query, value and key vectors, respectively, obtained through learned projections of the input representations while $d_k$ is the dimensionality of the query and key vectors. In summary, attention accomplishes selective learning by determining which features or connections to a particular particle in a data space are the most important for classifying and solving a particular task. Transformers are general ML pipelines utilising layers of attention. Readers are directed to \citet{Vaswani_2017} for further detail on attention and transformers. LitePT \citep{Yue_2025} is one such transformer, specialising in efficient computation of large particle groups. It may be applied to a 3D data cloud to compute learned features with attention. Depending on the size of the data cloud, the input would need to be partitioned in order to ensure the algorithm stays within computational limits. In section \ref{sec:ml_algorithms_benchmark_testing}, we will discuss the application of these methodologies to classify stellar streams.

\section{Benchmark Dataset}
\label{sec:benchmark_dataset}
\begin{figure*}
    \centering
    \includegraphics[width=0.16\linewidth, height=0.16\linewidth]{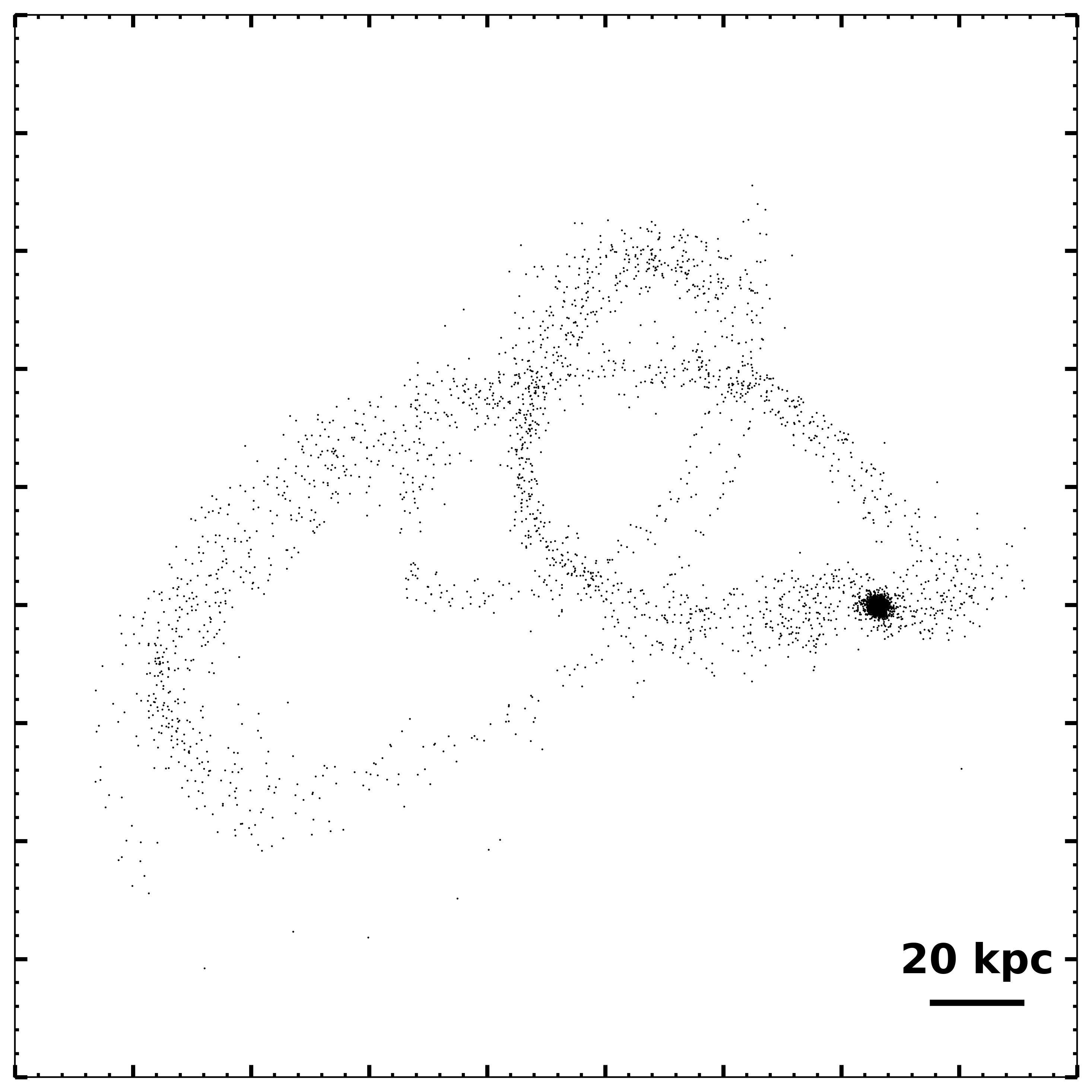}%
    \includegraphics[width=0.16\linewidth, height=0.16\linewidth]{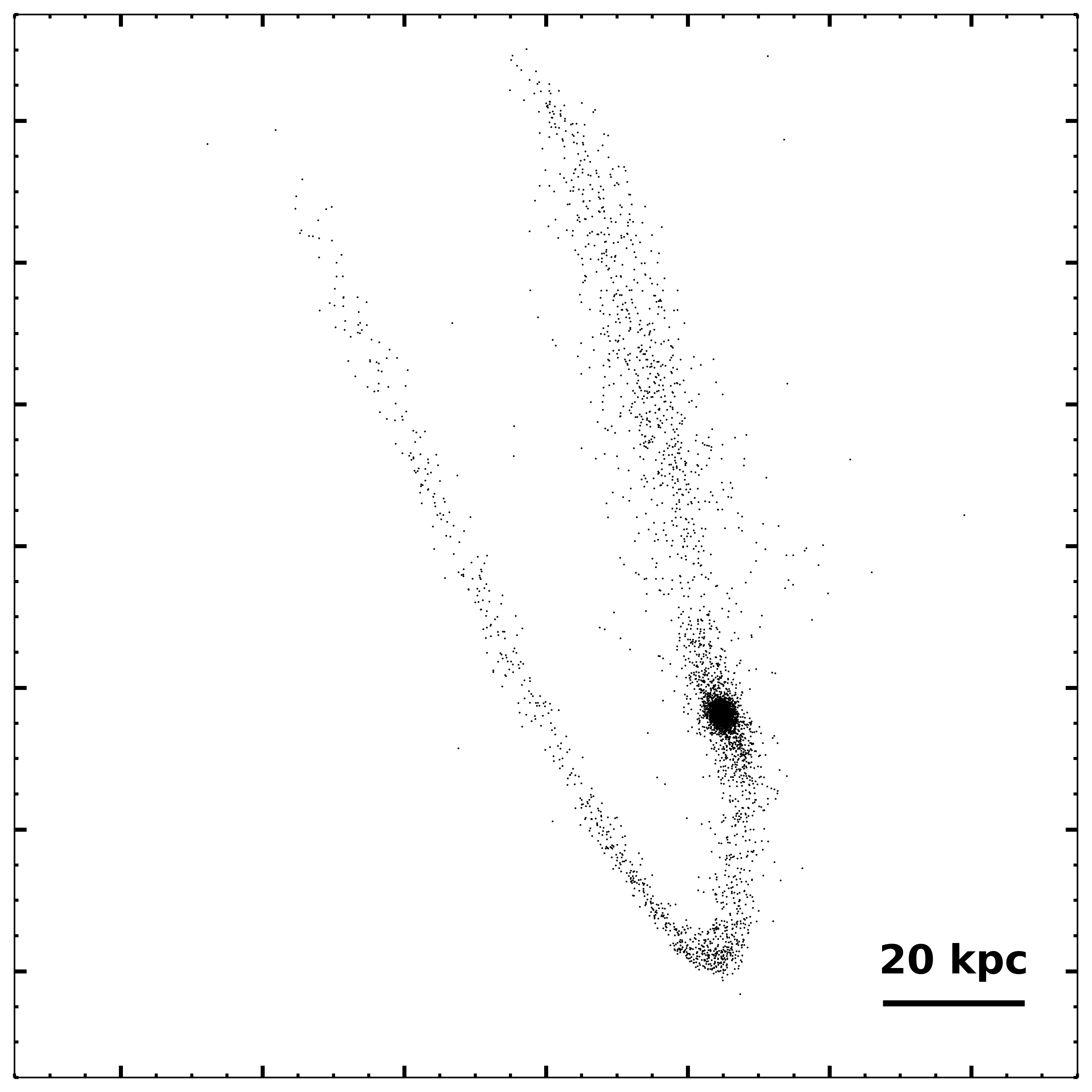}%
    \includegraphics[width=0.16\linewidth, height=0.16\linewidth]{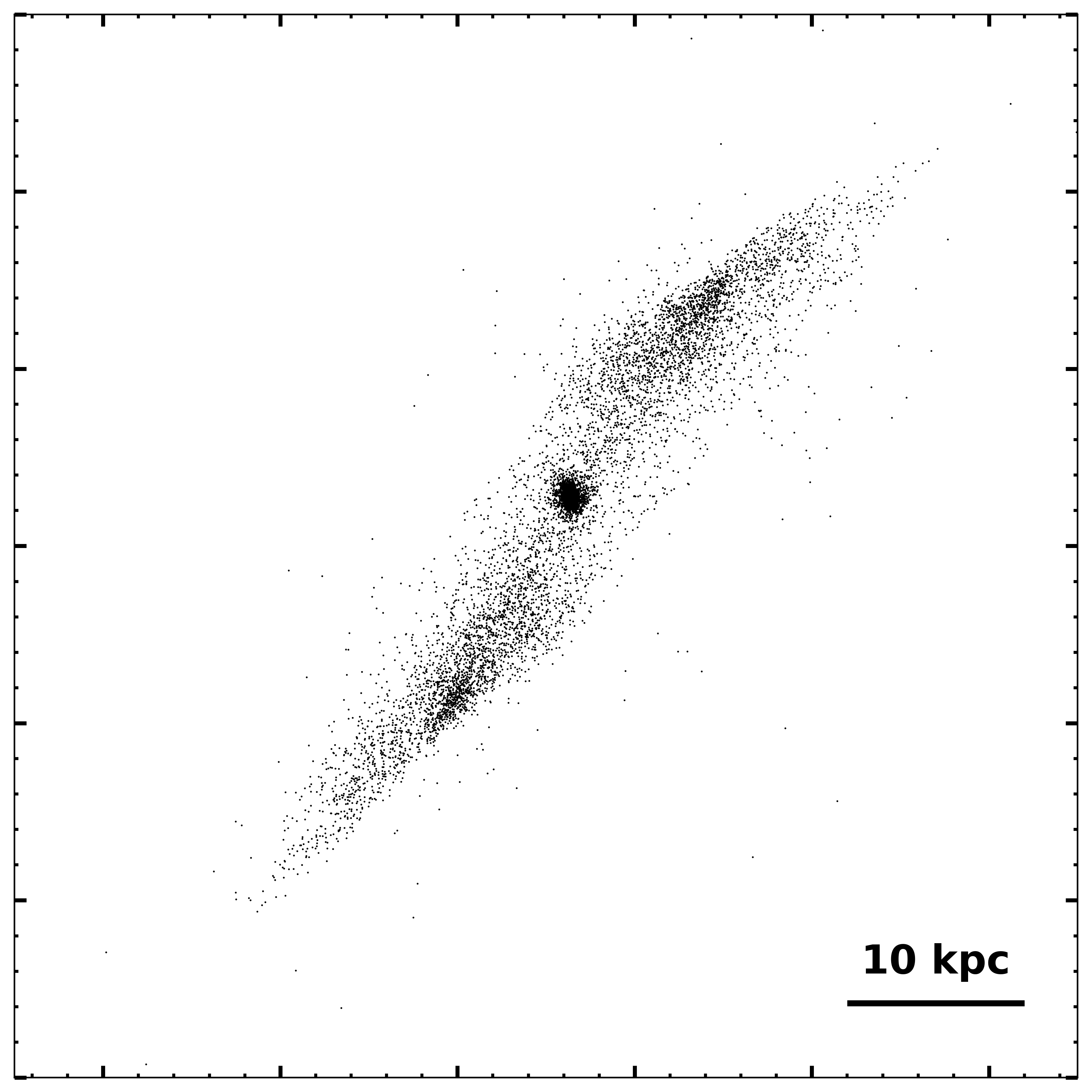}%
    \includegraphics[width=0.16\linewidth, height=0.16\linewidth]{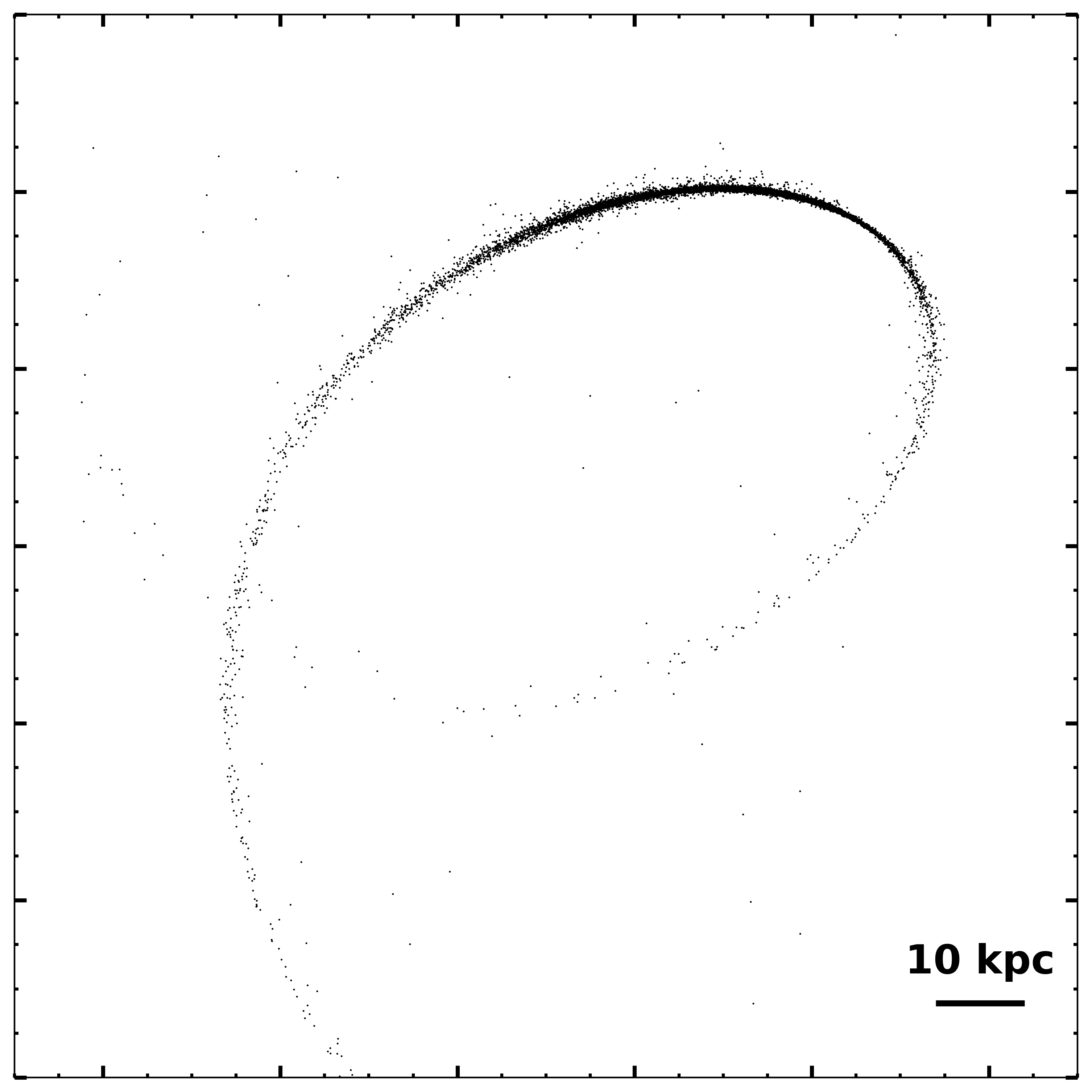}%
    \includegraphics[width=0.16\linewidth, height=0.16\linewidth]{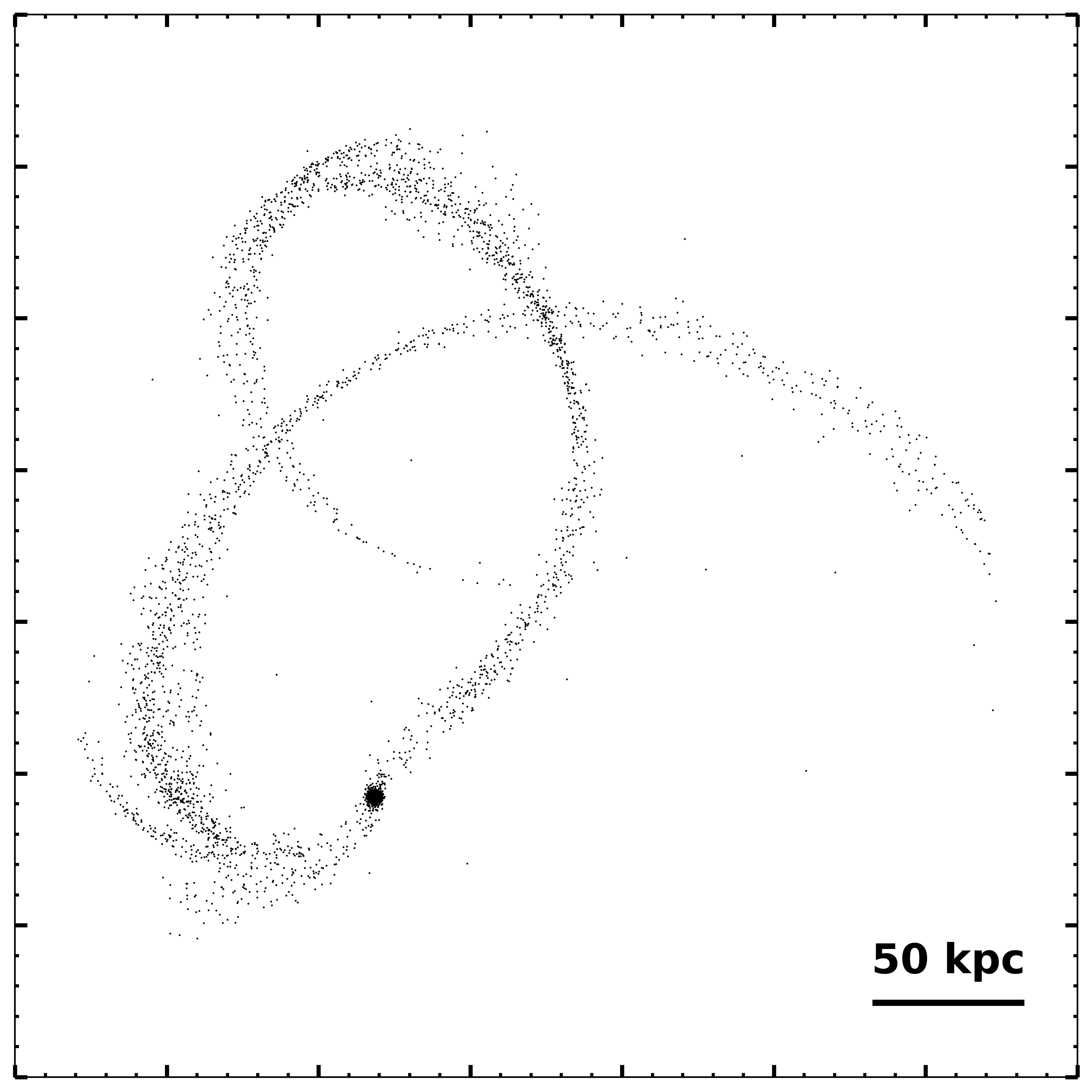}%
    \includegraphics[width=0.16\linewidth, height=0.16\linewidth]{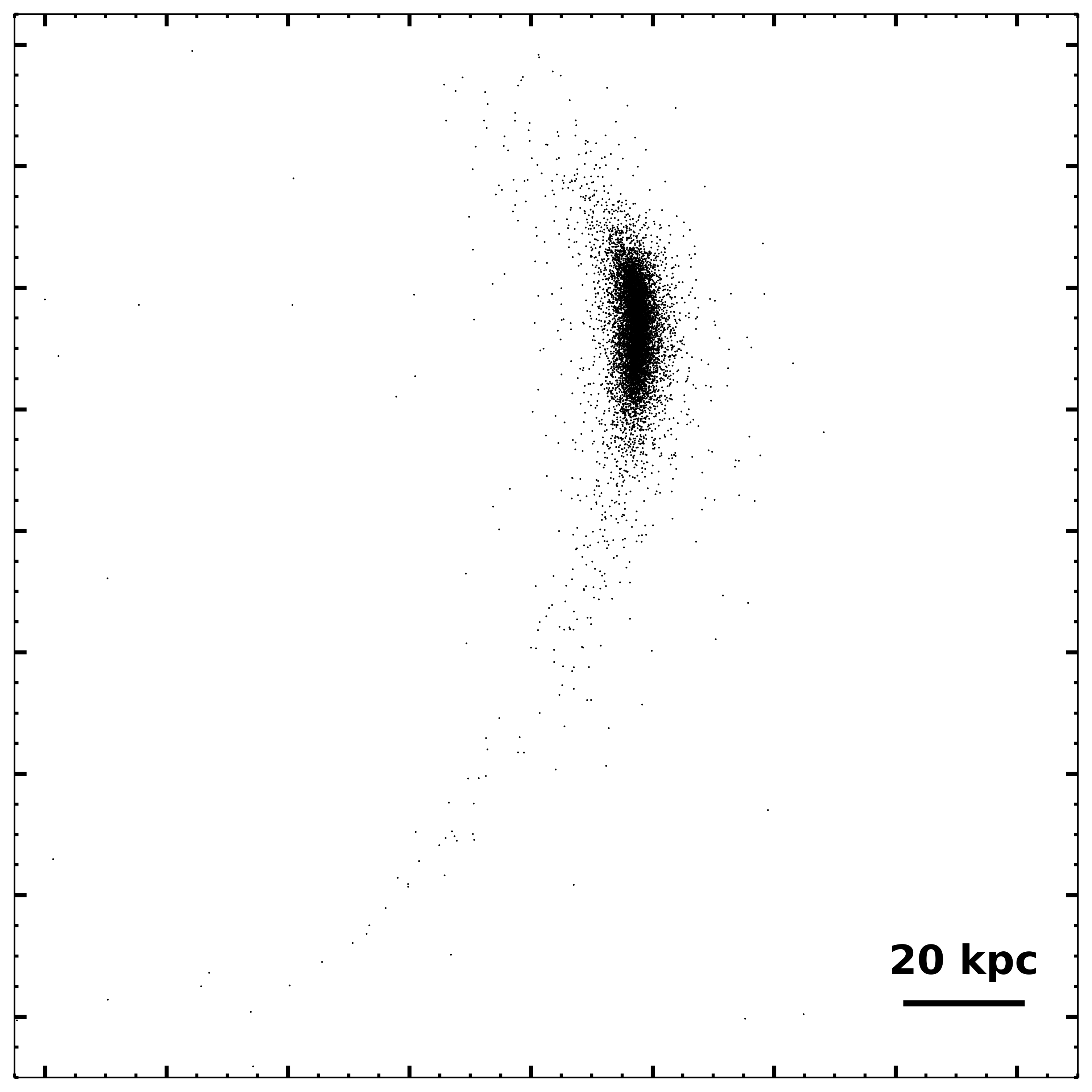}
    \includegraphics[width=0.16\linewidth, height=0.16\linewidth]{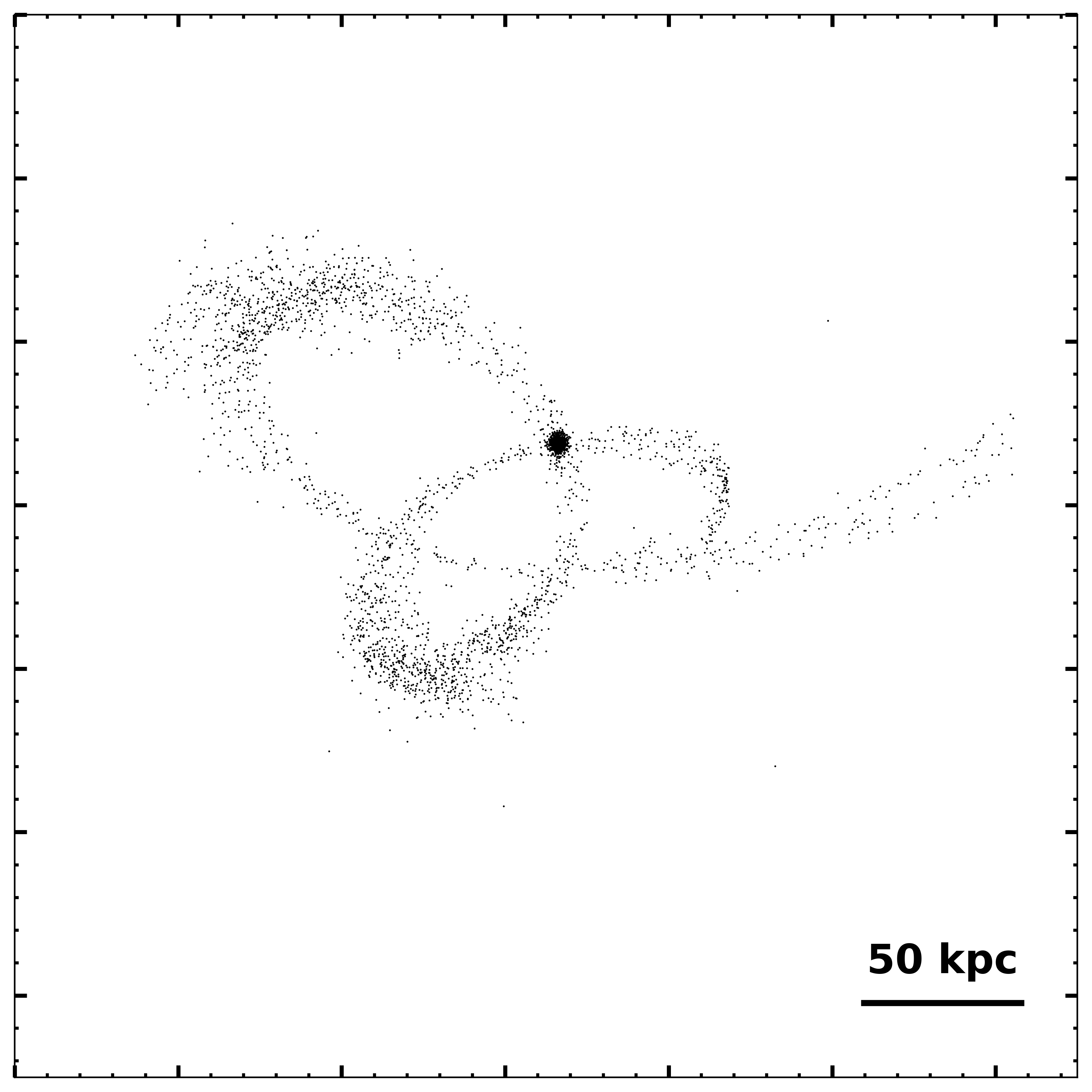}%
    \includegraphics[width=0.16\linewidth, height=0.16\linewidth]{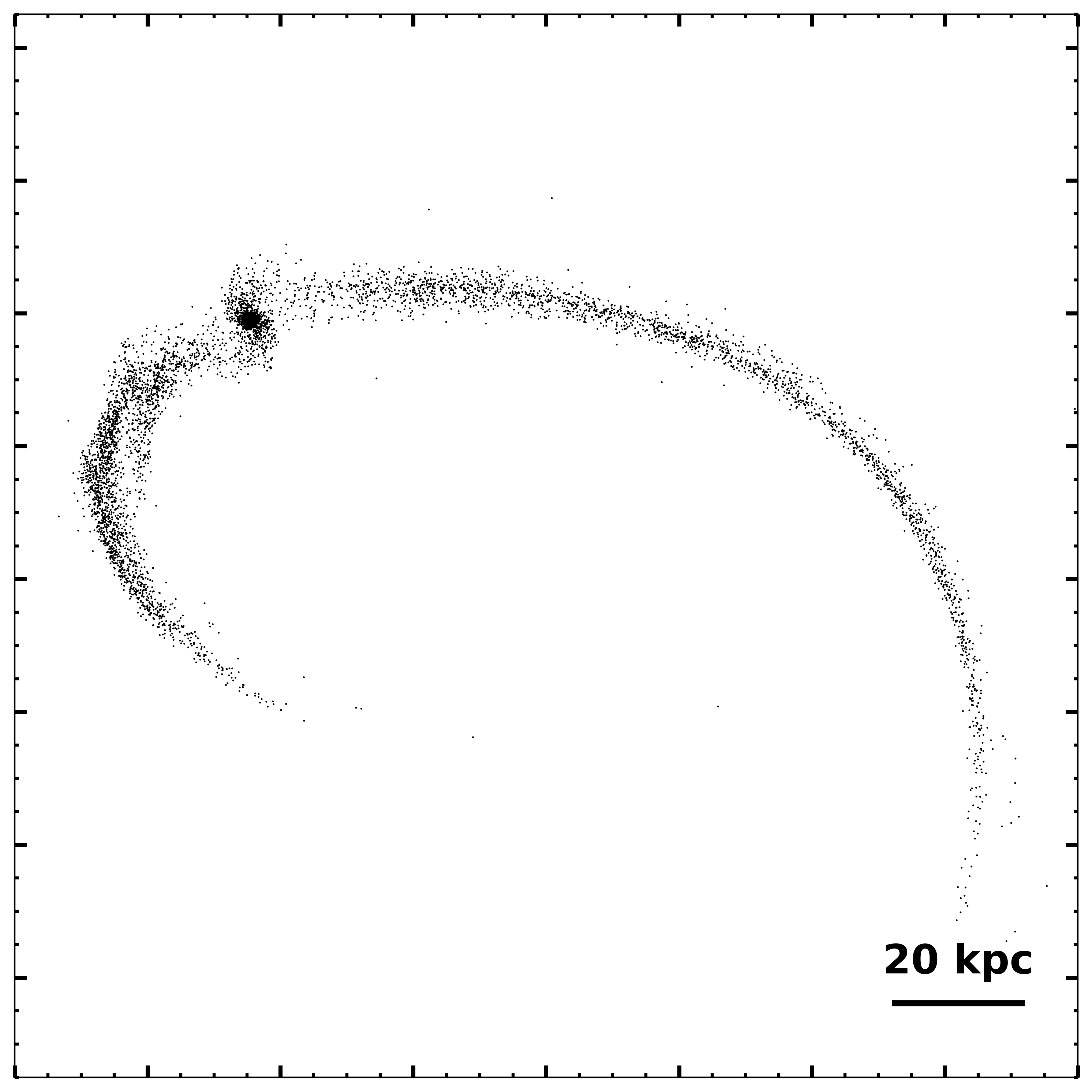}%
    \includegraphics[width=0.16\linewidth, height=0.16\linewidth]{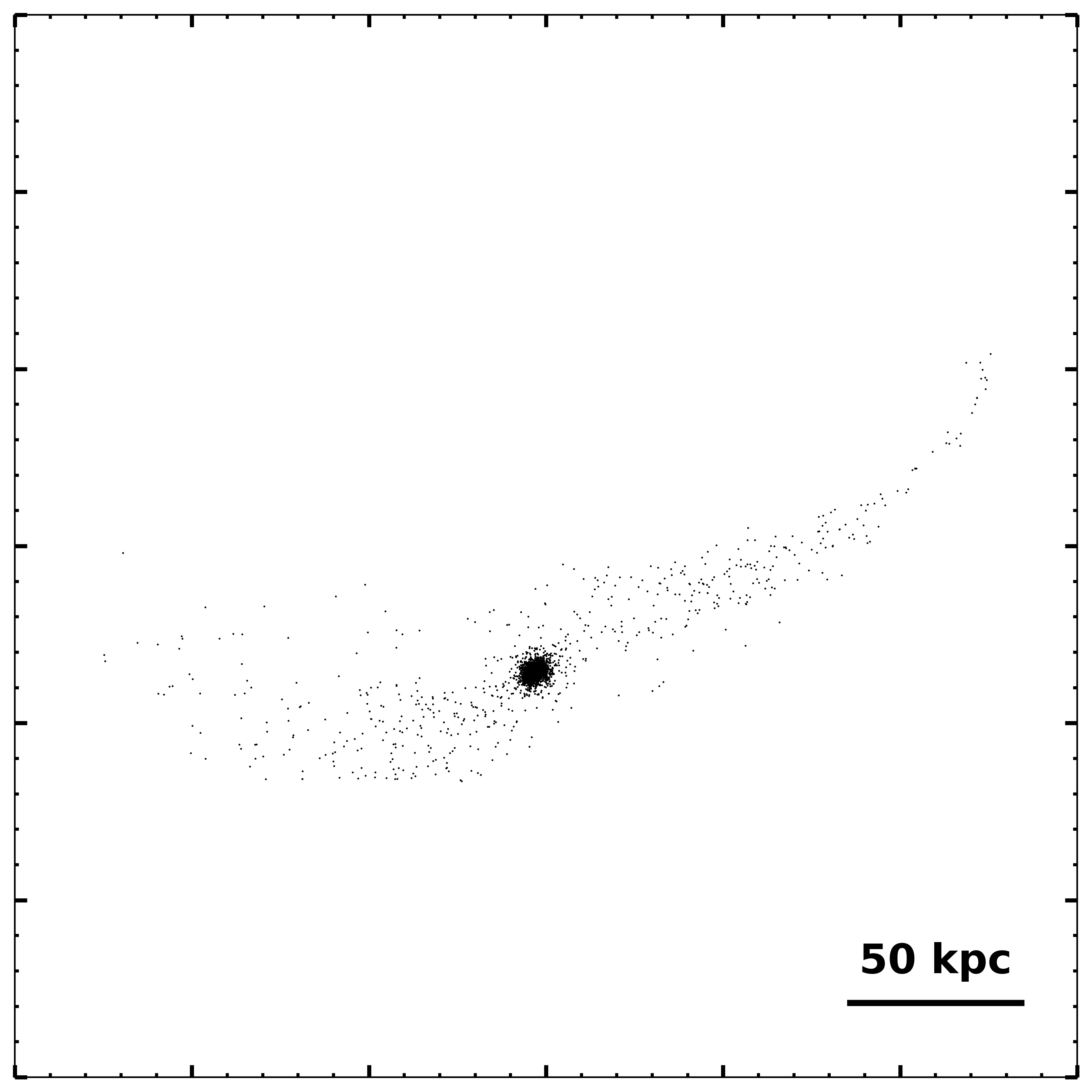}%
    \includegraphics[width=0.16\linewidth, height=0.16\linewidth]{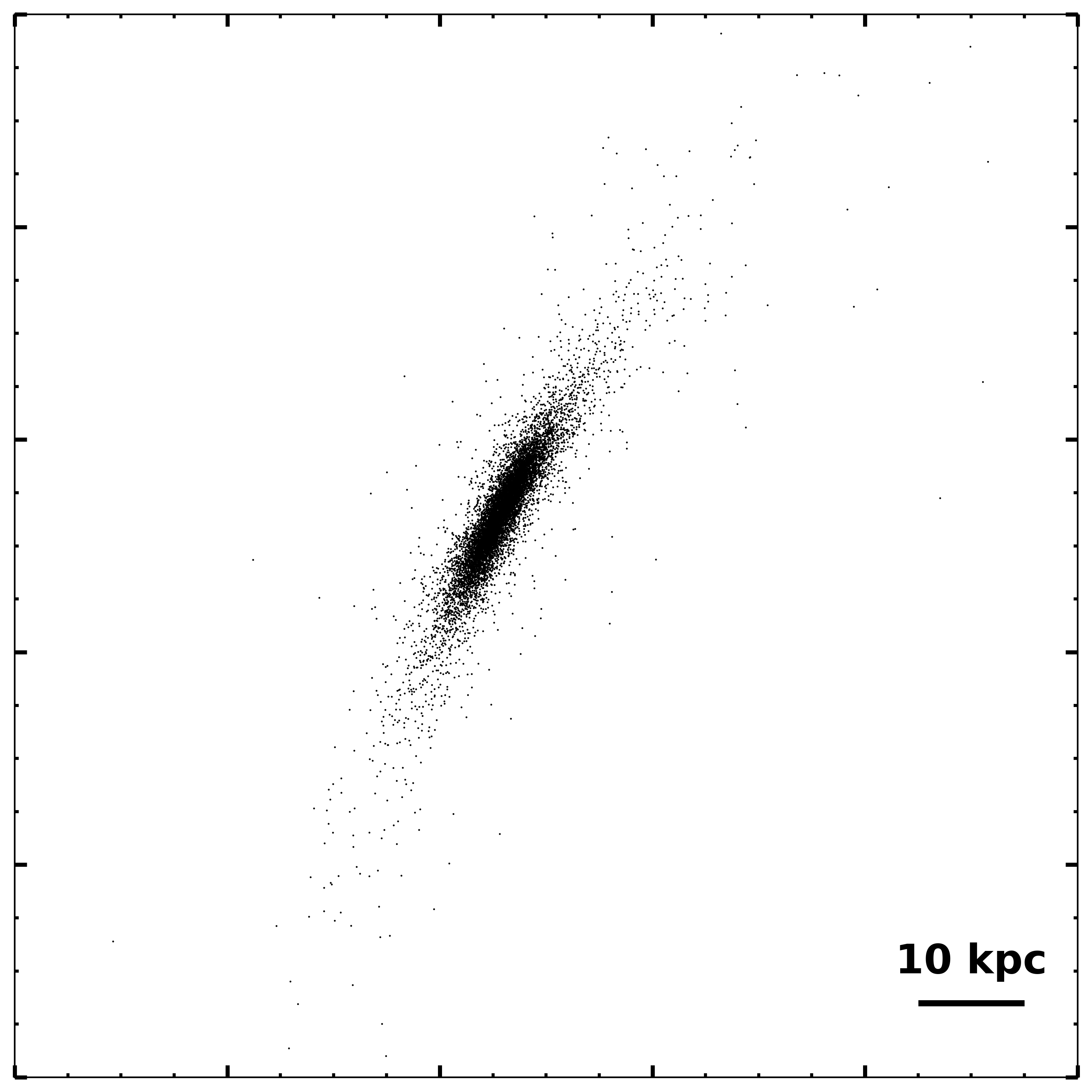}%
    \includegraphics[width=0.16\linewidth, height=0.16\linewidth]{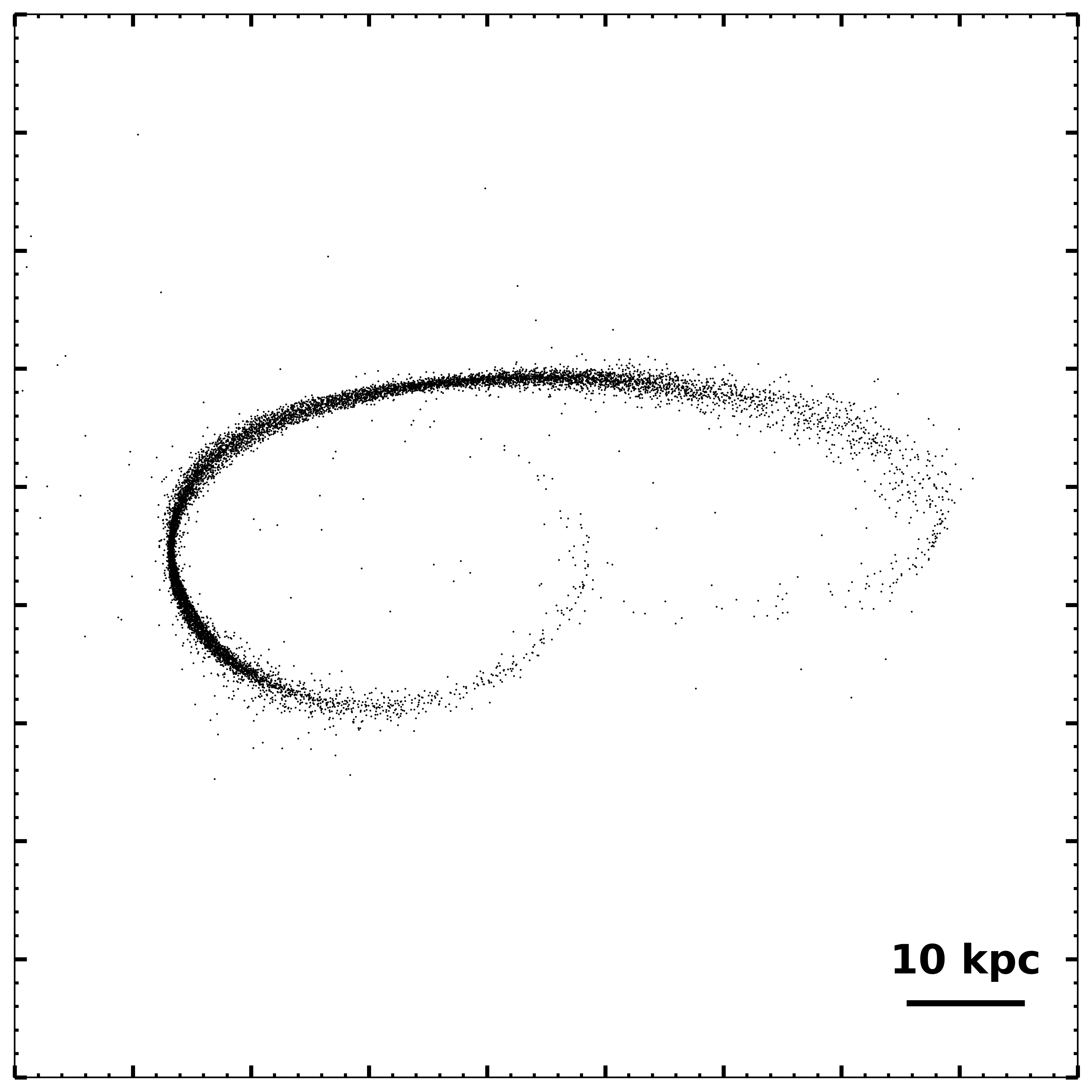}%
    \includegraphics[width=0.16\linewidth, height=0.16\linewidth]{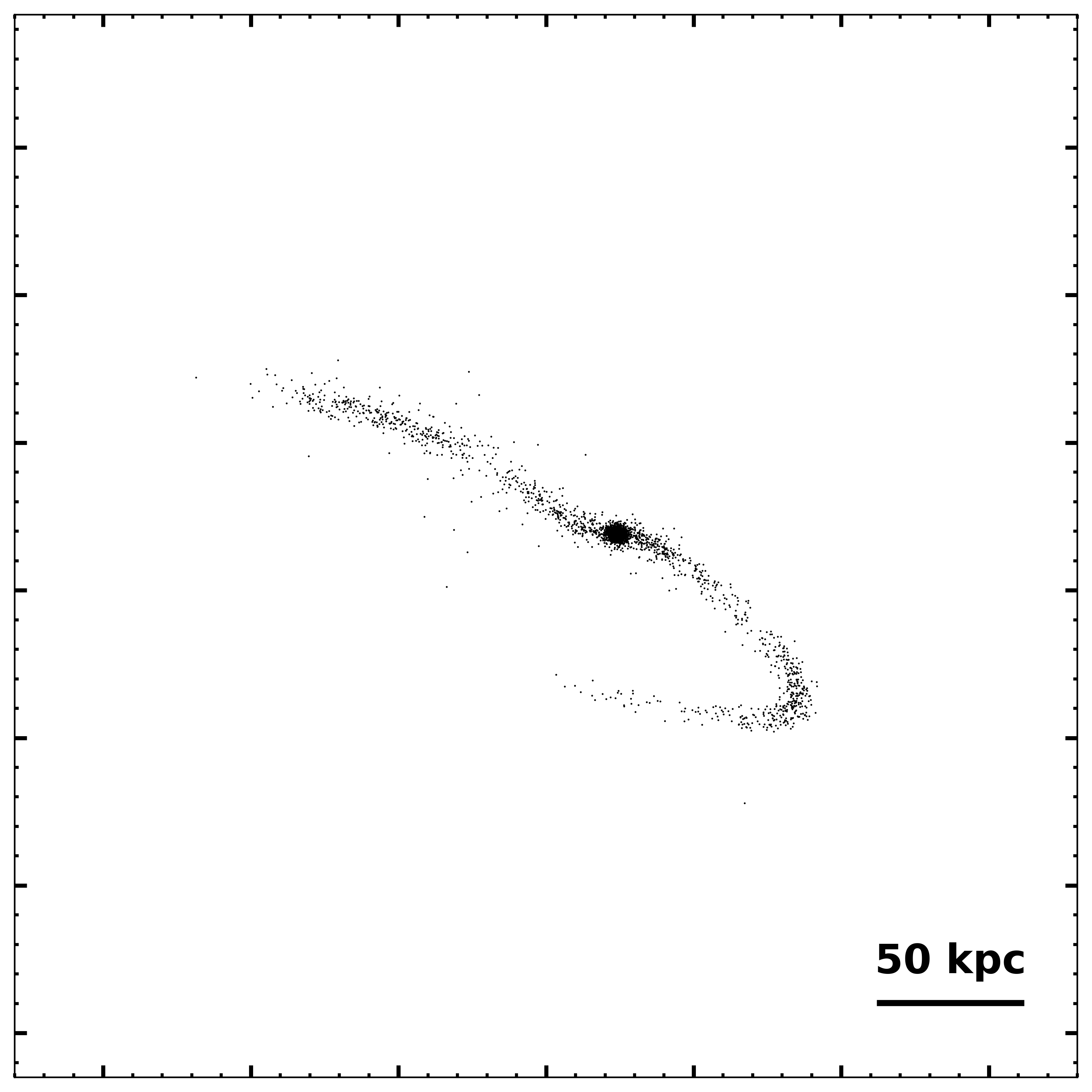}%
    \caption{A sample of stellar streams produced for the training, validation and testing datasets.}
    \label{fig:example_streams}
\end{figure*}

We construct our training and validation datasets with isolated, synthetic stream and background structures. Although particle spray methods \citep{Varghese_2011,Fardal_2015,Chen_2025} are popular, we instead chose to utilise Gadget-4 \citep{Springel_2021} based N-body simulations. All stellar streams were derived from Plummer sphere subhalos initialised as per \citet{Aarseth_1974}, with potential: 
\begin{equation}
    \Phi_{\text{Plummer}} = -\frac{GM_{\text{P}}}{\sqrt{r^2 +a^2}},
\end{equation}
We adopted a fixed Plummer scale radius of $a=1$ kpc across the progenitor mass range, without imposing a mass–radius correlation. Although this is not a realistic mass to size model, at fixed scale radius and comparable orbital conditions, lower mass progenitors are less dense and more susceptible to tidal disruption and vice versa. This produces the desired variation in the extent and morphology of their debris for training our ML algorithms.

Each progenitor was represented by $N$ particles with mass $\frac{M_{\text{P}}}{N}$. Most realisations contained approximately $10^4$ particles, with additional systems containing approximately $10^2$ to $10^5$ particles. The motivation here is to broaden the variations in the structures available for training the ML models, rather than reproducing a realistic satellite population or specific mass-size distribution. Furthermore, mass is not supplied explicitly to the classifiers, which instead use the resulting spatial and kinematic information. This is a simplified dataset created with the aim of creating a general suite of stellar streams. Thus it is worth acknowledging that alternative progenitor profiles and particle resolutions could change both the evolved debris and its representation in the training data. For viability in transfer across to independent data, we provide tests against the Auriga simulation suite.

These were placed into an external Milky Way-like flattened potential \citep{Smrithi_2026}, with a standard \citet{Hernquist_1990} bulge,
\begin{equation}
    \Phi_{\text{bulge}} = -\frac{GM_{\text{bulge}}}{r_{\text{bulge}} +r},
\end{equation}
with scale radius $r_{\text{bulge}}=0.7$ kpc and bulge mass $M_{\text{bulge}}=0.7\times10^{10}$ ${\rm M_\odot}$.

\citet{Miyamoto_Nagai_1975} disk, 
\begin{equation}
    \Phi_{\text{disk}}(R,z)=-\frac{GM_{\text{disk}}}{\left(R^2+\left(r_\text{disk}+\sqrt{\left(z^2+b^2\right)}\right)^2\right)^{1/2}},
\end{equation}
where scale height $b$ is set to a fifth of the disk radius. The scale radius is $r_{\text{disk}}=3.5$ kpc and the disk mass is set to $M_{\text{disk}}=0.7\times10^{10}$ ${\rm M_\odot}$.

and an NFW \citep{Navarro_1997} halo flattened as:
\begin{equation}
\Phi_{\text{halo}} = -\frac{GM_{\text{halo}}}{r} \ln\left(\frac{r}{r_{\text{halo}}}+1\right),
\end{equation}
where radius $r$ is,
\begin{equation}
    \sqrt{x^2 + y^2 + (z/q)^2},
\end{equation}
with flattening parameter $q$. Here, $r_{\text{halo}}$ is defined as $r_{\text{halo}}=\frac{r_{\text{vir}}}{c}$ with concentration parameter $c=10$, and virial radius $r_{\text{vir}}=282$ kpc per \citet{Bland-Hawthorn_2016}.

To further promote variation, we randomised initial conditions with bounds listed in table \ref{tab:train_val_ranges}. The advantage here is that stellar stream formation is not guaranteed. Along with increasing the variability of formed streams, the dataset also provides a valuable negative, contributing to the background (non-stream) dataset. The negative data include subhalos in very early stages of disruption as well as coherent non-stream-like structures. These included shells, plumes, partially and fully phase mixed structures. The motivation here is to inform the ML algorithm that streams are not just generic coherent structures, but rather a specific subset. This naturally leads to the question of how to define a stellar stream. 

Taking into consideration the factors discussed in section \ref{sec:background}, for the work presented here, we primarily classify stellar streams from a morphological perspective. Orbital history acts as a secondary delineator. The classification pipeline flowed as follows: First, we randomise the initial conditions of the subhalo progenitors based on typically observed globular cluster and dwarf galaxy based stellar streams \citep[and the references within]{Mateu_2023}. We take the mass, $\log_{10}(\mathrm{M}_\odot)$, randomised between $5-10$. Initial position was determined by first selecting an initial radial distance $10-300$ kpc from the centre of the host halo potential. From there, the $[x,y,z]$ components were randomly sampled. A similar process was followed for initial velocity with bounds of $0-250$ km/s. These were injected into the flattened potential with $q=0.85, 0.9, 1, 1.1, 1.2$ to account for prolate, spherical and oblate halo scenarios. All subhalos were evolved for a period of $10$ Gyrs. However, to capture the disrupted structures in various stages of evolution, the chosen structures are from a range between $1-10$ Gyrs. For background, some slightly disturbed halos $<1$ Gyr were also selected. 

\begin{table}
	\centering
	\caption{Range of values used to initialise the evolving subhalos for the training and validation datasets.}
	\label{tab:train_val_ranges}
	\begin{tabular}{lll} % four columns, alignment for each
		\hline
		Parameter & Minimum & Maximum \\
		\hline
		Mass $(M_{\odot})$ & $10^5$ & $10^{10}$ \\
		Initial (Radial) Position (kpc) & 10 & 300 \\
		Initial Velocity Magnitude (km/s)  & 0 & 250 \\
        Time (Gyr)  & 1 & 10 \\
        Flattening (q)  & 0.85 & 1.2 \\
		\hline
	\end{tabular}
\end{table}

The stream classification process was conducted with visual inspection. Primary markers of streams considered were the characteristic arms and orbital history. Fig. \ref{fig:example_streams} provides a brief sample of chosen streams. Noting, however, that this is non-exhaustive, and a comprehensive variety of stream morphologies were selected. It is worthwhile acknowledging that with visual inspection, the definition and boundary of what was considered a stream was completely morphological and thus to a degree, arbitrary. Although streams with diffuse tails were selected, any structure with fanning, resembling shell-like structures were rejected. In these cases orbital history was strongly considered, with structures formed with (near) radial orbits with centre pass throughs excluded. In contrast, those with diffuse tails in less eccentric orbits were accepted. However, if the tails possessed excessive concentrations of unbound particles which no longer followed the orbital path of the main body, they too, were rejected. Additionally, thin long streak-like structures in the intermediary phase of centre pass through in shell formation were also excluded despite their similarity to stream-like structures. Other rejected samples included: plume-like stream-shell hybrid structures, high mass subhalos with minimal disruption with pericentres at large radii, other minimally disturbed subhalos that drift away or have orbital periods well in excess of the $10$ Gyr evolution period and a variety of other phase-mixed or non-distinct structures. Some stream samples were slightly trimmed of unbound particles drifting tens of kpc away from the main body. 

At the conclusion of the classification process, we obtained $2184$ stream samples from approximately $3300$ simulations. In the case of ML, typically, the larger the dataset, the greater the performance. Thus, there exists no upper cap for any particular problem. However, this is not without other considerations. In particular, increasing the size of the dataset also results in an increased computational cost. For the work here, we remain at this range as we find strong results, which will be discussed later. However, the growth of the dataset is considered ongoing work. The total stream sample was approximately divided into a 6.5:1:1, training to validation to testing ratio. This was chosen to maximise training variety while maintaining appropriate samples for validation and testing. We maintain this ratio for all ML methodologies for consistency. 

The background data was composed of a collection of separate features. The primary object included was a Milky Way-like analogue. The training data utilised a typical Hernquist, Miyamoto-Nagai, NFW profile. The validation employed a similar Milky Way analogue, with slight variation to ensure discernible difference from the training halo. The testing set, however, contained an exponential disk \citep{Freeman_1970}, Jaffe bulge \citep{Jaffe_1983} and Einasto halo \citep{Einasto_1965} to form a similar Milky Way-like structure \citep{Widrow_2008,McMillan_2017}. The motivation for the different profiles was to ensure an accurate depiction of stream detection accuracy during validation and testing. The use of the same distribution would provide a trivial negative. In addition to the galactic structures, the background was supplemented with a collection of undisturbed or very minimally disturbed subhalos of varying mass, position and velocities. This distribution was chosen randomly. On top of this, a sample of the rejected non-stream-like structures from the classification dataset was also injected. The motivation here was to introduce maximal background noise. The addition of similar structures from the classification dataset also provided the presence of additional coherent structures with similar position and velocity distributions. The aim was to ultimately ensure the ML algorithms learn to define a stream, rather than memorise the collection of training data provided.

\section{ML Algorithms and Benchmark Testing}
\label{sec:ml_algorithms_benchmark_testing}
Next, we define and apply our four ML frameworks to the dataset. To quantify their effectiveness, we run a benchmark test on two scenarios. First, with a halo only background. Here we produce an environment containing only the streams and the test Milky Way-like halo previously discussed. Secondly, we increase the complexity by introducing other non-stream, coherent structures similar to training and validation. With the exception of XGBoost, the ML techniques all utilise Matthews Correlation Coefficient \citep[MCC;][]{Matthews_1975},
\begin{equation}
    \mathrm{MCC} =\frac{TP\times TN-FP\times FN}{\sqrt{(TP+FP)(TP+FN)(TN+FP)(TN+FN)}}
\end{equation}
to train and select the best model. Here, True Positives (TP), True Negatives (TN), False Positives (FP) and False Negatives (FN) are weighted to provide a measure of accuracy, precision and recall. It ranges from $-1\leq\mathrm{MCC}\leq1$, where $+1$ indicates perfect classification and $-1$ indicates perfect disagreement. A score of $0$ essentially corresponds to random classification. This was chosen due to its resilience against class imbalance problems such as our case here where we expect stream particles to be a small fraction of a halo's total particle count. For a more intuitive assessment, we also evaluate using several other performance metrics. Namely; precision, the fraction of predicted stream particles that are true stream members and recall, the fraction of true stream particles successfully recovered. In addition, there are the performance metrics F1, a weighted mean of precision and recall as well as intersection over union (IoU), the number of correctly identified stream particles divided by the number belonging to either the true or predicted stream population. Given the nature of the problem, in particular the class imbalance between stream and non-stream members, MCC was chosen over F1 and IoU as the primary performance metric. It is also considered a superior measure to F1 in binary classification models \citep{Chicco_2020}. During inference, the chosen best model is applied to a complete testing dataset and MCC was once again used as the primary metric to gauge their performance. All four codes are provided in the stellar stream semantic segmentation GitHub \footnote{https://github.com/Viraj-E/ML-stellar-stream-classification} page.

For the models to be discussed, the input features were chosen to describe the spatial morphology and kinematic coherence of the stream population defined in section \ref{sec:benchmark_dataset}. Neighbour distances and spatial covariance characterise local concentration and elongation, while velocity differences, dispersions and angular momentum alignment provide further kinematic context. Host centred quantities provide additional galactic context. The motivation for these quantities is to provide potentially informative descriptors of spatial and kinematic structure, allowing the algorithms to learn which combinations distinguish the morphologically selected streams from the background population. We do not impose a predetermined physical classification rule or optimise for any specific definition of a stream. Instead, the decision boundaries are learned empirically from the labelled examples. The adopted features constitute a physically motivated representation whose usefulness is assessed through the reported classification performance. It is important to acknowledge that ML performance depends on the model, input features, training procedure and data, and must be assessed empirically. Our aim is therefore to establish the comparative performance of the adopted pipelines, rather than demonstrate global optimality through an exhaustive search of these interacting choices.

\subsection{XGBoost: Feature Decision Tree}
As previously discussed, XGBoost is not, strictly speaking, a deep learning algorithm. However, it can prove effective at finding the weight, or importance of user defined features. In terms of information input, we restrict ourselves to exclusively phase space (3D positional and 3D velocity) data. Starting the process, we first construct a tabulated list of a sample data space containing a collection of stream and background particles. For the work presented in this paper, this equated to approximately 300,000 stream and 3 million background particles in the training dataset. We chose the 1:10 ratio to reflect the class imbalance that would typically be present in realistic halos. That is, we are informing the algorithm to expect environments more rich in background and non-stream particles. Considering the creation process, the stream and background populations are dynamically disconnected. However, they exist in a similar kinematic regime, providing an approximate imitation of a realistic environment.  

With the dataset at hand, we must determine an appropriate metric to separate stream-like from background particles. For this purpose, raw position and velocity coordinates by themselves are a weak choice. This is because it teaches the algorithm that specific position or velocity values are important indicators of a stream. Considering all the training data is synthetic and evolved against an arbitrary axis orientation, any learned weightings would translate poorly to independent testing data. This effect is further compounded by the fact that XGBoost is unable to learn new features and thus must rely solely on the provided input values. Thus, the natural next step is building an effective feature vector $\zeta$ that is able to navigate the problem at hand. We can take advantage of the available phase space vector for each particle by computing derived kinematic quantities that are more effective at separating stream-like from background particles. For a comprehensive feature representation, we take a combination of individual and local group statistics. In the case of individual values, for each particle $[\mathbf{x}_i,\mathbf{v}_i]$, we compute the radius $r_i$, the speed, $|\mathbf{v}_i|$, the radial $v_{i,r}$ and tangential velocities $v_{i,t}$ and angular momentum magnitude $|\mathbf{L}_i|$. Unless otherwise stated, particle positions and velocities are expressed relative to the host centre. Similarly, radii, radial and tangential velocities, and specific angular momenta are calculated with respect to the centre of the host halo. This applies to all four ML models discussed.

The k-nearest neighbours (kNN), the k particles closest to a selected particle under the adopted distance metric, provide the remaining quantities to create the nearest neighbour vector $\mathcal{N}$. This process involves selecting a particle and defining it as the centre, $\mathbf{c}_i=[\mathbf{x}_i,\mathbf{v}_i]$. From here, a kNN patch is constructed around the centre particle. Here, we take a patch of $N=48$. This value was chosen to optimise the balance between capturing local versus global structure. A lower value may be too localised and sensitive, not capturing a general enough depiction of a stream or background local environment. On the opposite end, a larger $N$ value runs the risk of excessive noise as larger patches with stream and background particles will be less coherent. In addition, it increases computing and memory costs. Once the patch is formed, we compute a large selection of values to maximise the information we provide the algorithm. This is especially important as we continue to keep in mind that XGBoost cannot learn new features and only relies on what information we provide.

For these local group features, first, we consider a density perspective. Stellar streams can typically be considered local regions of overdensity. This generally holds true at radii $>\sim10$ kpc, away from the excess stellar matter of the galactic disk. The density may be characterised in several ways, including computing local number or mass spherical densities extending from the centre particle. However, to reduce computational cost and to take advantage of the constructed kNN patch, we instead take two density analogues. For each particle, we compute the average distance from its neighbours as $\bar{d}_i=1/k\sum_j|\mathbf{x}_i-\mathbf{x}_j|$ as well as the distance to the furthest neighbour $d_{i,max}$. In addition to this, we also input information on the directionality of the density via two spatial dispersion features. These are provided by the eigenvalues of the covariance matrix
\begin{equation}
    C=\frac{1}{k}\sum\mathbf{x}_j'\mathbf{x}_j'^{\mathrm{T}},
\end{equation}
where $\mathbf{x}_j'=\mathbf{x}_j-\bar{\mathbf{x}}$. This provides eigenvalues $\lambda_1\geq\lambda_2\geq\lambda_3$. We take the ratios $\alpha_{i,1}=\lambda_1/\lambda_3$ and $\alpha_{i,2}=\lambda_2/\lambda_3$. From a velocity perspective, we also take the standard and radial velocity dispersions $\sigma_{i,v}$ and $\sigma_{i,v_r}$. Contributing further, we have mean speed ratio $s_i=1/k\sum_j|\mathbf{v}_j|/|\mathbf{v}_i|$, mean tangential velocity $\bar{v_{i,t}}=1/k\sum_j v_{j,t}$ and mean relative radial velocity $\bar{v'_{i,r}}=1/k\sum_j|v_{j,r}-v_{i,r}|$. These quantities were chosen as streams typically tend to possess velocities aligned along their orbital path and possess low variance radially. 

The final feature utilised is angular momentum alignment. Similar to speed, we compute the average angular momentum ratio $R_i=1/k\sum_j|\mathbf{L}_j|/|\mathbf{L}_i|$. For each patch, we take $\hat{\mathbf{L}}_j$, and calculate the mean angular momentum alignment $\mathbf{A}_i=1/k\sum_{j}\hat{\mathbf{L}}_j$. From this we obtain the magnitude $A_i=|\mathbf{A}_i|$. Here, we are taking advantage of the fact that stellar streams are typically aligned in angular momentum space, and thus stream particles should possess $A_i>0$ while background centres will have $A_i\approx0$. 
Collecting these properties together gives us, the nearest neighbour vector
\begin{equation}
    \mathcal{N}_i=[\bar{d}_i,d_{i,max},\alpha_{i,1}, \alpha_{i,2},\sigma_{i,v},\sigma_{i,v_r},s_i, \bar{v_{i,t}}, \bar{v'_{i,r}},R_i, A_i]
\end{equation}
Thus, we obtain the input feature vector for the XGBoost algorithm,
\begin{equation}
    \zeta_{i,XGB}=[r_i,|\mathbf{v}_i|,v_{i,r},v_{i,t},|\mathbf{L}_i|,\mathcal{N}_i].
    \label{eqn:feature_vector_xgboost}
\end{equation}
Importantly, we exclude the raw phase space $[\mathbf{x}_i,\mathbf{v}_i]$ information from the input to prevent biased learning by the algorithm. As has been previously discussed, the user input is the only information that the decision tree is able to utilise to characterise stream particles. Adding phase space values brings the risk of bias to absolute position or velocity values. We do not want the algorithm to gain preferences for certain orientations or $x-y-z$ component position and velocity ranges. This would lead to false positives in galactic systems where many non-stream particles can occupy a very similar phase space. 

With the input feature vector finalised, we are able to initialise XGBoost. Following the gradient boosting formalism of Eqn. \ref{eqn:gradient_boosting}, the algorithm proposes a binary decision tree utilising a quantity of the input feature. The particles are categorised and a loss value as per Eqn. \ref{eqn:gradient_boosting_loss} is computed. This process is iterated with further trees sequentially added to minimise the loss function. The end result is an internal decision tree $\Phi$ with $n$ trees that classify stream-like particles. This creates a stream `likeness' score that ranks how stream-like a certain particle is. This can be represented as a probability analogue stream-likeness score for particle $i$, as:
\begin{equation}
    y_{i,xgb}=F_{XGB}\left(w_n\Phi_n(\zeta_{i,XGB})\right).
\end{equation}
Here, the XGBoost function, $F_{XGB}$ produces the ranking score as a function of each decision tree applied to the feature vector, $\Phi_n(\zeta_i)$. Not all trees are considered equal, with each tree assigned a weight $w_n$ depending on their effectiveness in separating stream from background particles. Unlike the deep learning methods hereafter, the XGBoost code uses Area Under Curve \citep[AUC;][]{Hanley_1982} for model selection due to it being a more appropriate metric to measure ranking models. However, for testing results, we present the MCC score for consistency in model comparison. 

\begin{figure}
    \centering
    \includegraphics[width=\linewidth]{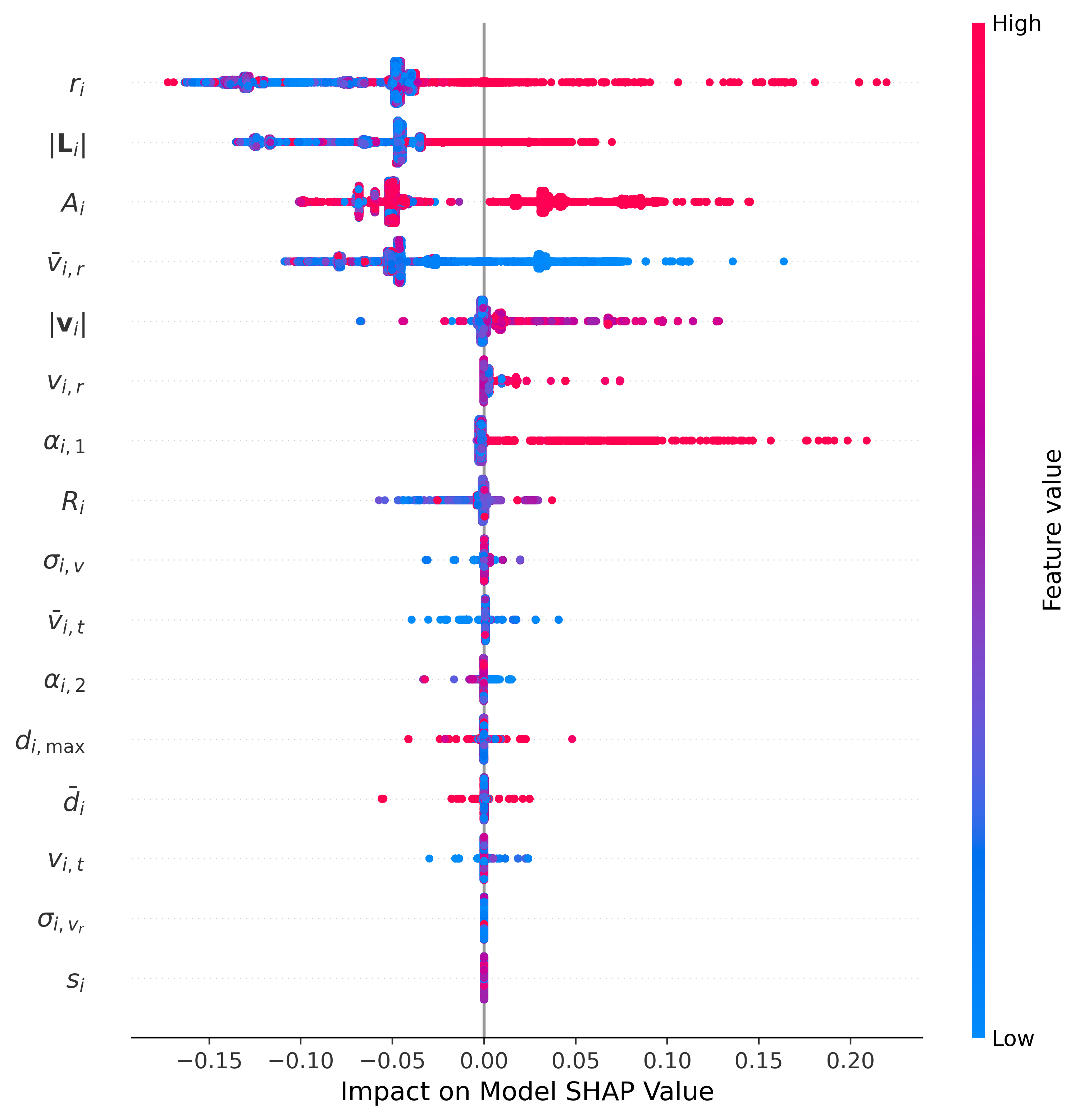}
    \caption{SHAP plot of features ordered by their contribution to classification. Each point represents a sampled particle.  Feature value is simply a representation of the magnitude of the value. This is displayed for each value in the XGBoost feature vector in order of importance to the model's internal classification metric. A more positive SHAP value corresponds to a higher stream-likeness score and vice versa. }
    \label{fig:xgboost_beeswarm}
\end{figure}

Inspecting the model's performance during testing, we are able to examine which features in Eqn. \ref{eqn:feature_vector_xgboost} the internal gradient boosting algorithm utilises to provide its classification. This is detailed in Fig. \ref{fig:xgboost_beeswarm}. The features are listed in order of significance in determining the XGBoost stream-likeness score. In other words, the higher the feature is listed, the more `important' it was considered to be in determining if a particle was stream-like. Each particle's vector quantity magnitude is ascribed into the feature value. The separation of higher and lower feature values indicates how the model used that value to classify a stream-like particle. We see that, for example, particles with higher radii, $r_i$, tend to possess a higher SHAP value (stream-likeness score). On the other end, this indicates that the model is disfavouring particles at a lower radii. This is ideal as it shows that the model is filtering out noise from the stellar disk and bulge. Physically this aligns with typical cold streams which are found away from the core of a galaxy. 

Moving on from radii, we see that the next most important features are the magnitude of angular momentum and angular momentum alignment, $\mathbf{L}_i$ and $A_i$ respectively. Once again, we see higher values tend to indicate stream-like particles with a higher SHAP value. Especially in the case of alignment, this is what we would expect to see from cold streams that have low variance in orbital phase space. This is further reinforced by the spatial dispersion measure $\alpha_{i,1}$. Interestingly, velocity dispersion, typically considered a strong measure of stream-like structures, falls into the middle of the priority rankings. The separated radial velocity dispersion is given an even lower weighting. The top three features all involving measures of position indicates that the XGBoost algorithm primarily utilised positional arguments in its determination of stream-like structures. However, given the algorithm's inability to learn new features, this may be a consequence of the input variables, rather than a natural extraction from the algorithm training process. 

\subsection{Voxel Segmentation}

Although typically applied across image or photometric data, here, we take a more novel approach to CNN, applying it instead across phase space. To accomplish this, we once again first generate a training data space which will then later be voxelised. Taking the same stream and background dataset, we apply a different methodology to the one used in the XGBoost code. We run into the important consideration of computational and memory costs of applying a CNN across a large dataset. To ensure viability, we instead utilise a sampled data space during training. To do this, we simply take randomised portions of particles from each stream and background data file and construct a synthetic data cloud. Once again, it is worth acknowledging that although it is unphysical, it still remains an effective training environment. From here, a random particle is selected as the centre and a spatial box is constructed around it. With resolution and computational costs in mind, a $128$ kpc box was selected for simulation. This box is then voxelised into a $32\times32\times32$ grid. A feature vector is then produced as an input for the ML algorithm. Importantly, the feature vector is for the entire voxel, as opposed to an individual particle. This is necessary as the CNN layer is applied across voxels rather than particles in the created box. 

When the resulting grid is processed, the 3D CNN layers apply shared local kernels throughout the grid, allowing the model to identify spatially coherent patterns in density and kinematics while progressively combining information from neighbouring voxels. The network compresses these learned responses into a 48D latent representation, which is passed through a final classification layer to produce stream and background logits. From this, the model assigns a class to each occupied voxel based on both its internal summary features and the surrounding 3D structure. During inference, overlapping boxes are evaluated and their predictions combined to reduce boundary effects.

As a voxel is a representation of all the particles within, the feature vector constructed is an effective summary vector for the entire voxel. The first value represented is density. Since the size of the voxel is constant, $\rho\propto N_{{\rm voxel}}$, we simply take the particle count as the density. However, since the data space is randomly distributed, the number of particles in a particular voxel can range from potentially $N_{voxel}\sim10-10^4$ or more. To prevent instability arising from such a large variance, we take density as log particle count: $\rho_V=\log(N_V+1)$. In addition, we include the average $x-y-z$ velocity and angular momentum components as well as averaged radial distance, radial velocity, tangential velocity and angular momentum magnitude. This provides us with the Voxel feature vector,
\begin{equation}
    \zeta_V=[\rho_V,\bar{\mathbf{v}},\bar{r},\bar{v}_r,\bar{v}_t,\bar{\mathbf{L}},\bar{|\mathbf{L}|}].
    \label{eqn:feature_vector_voxel}
\end{equation}
The physical quantities are calculated for individual particles with respect to the host centre before averaging. The voxel centre determines spatial binning and is not the origin used to calculate radial velocities or angular momenta. Notably, no 3D spatial coordinates are provided in the feature vector. This is due to the positional information already being encoded as the relative position of the voxel within the grid. However, radial position is provided as an extra data point to provide perspective with respect to the centre of the galactic halo. This is further supplemented with the angular momentum and circular velocity components. Finally, the stream and background logits are then utilised to compute the Voxel probability,
\begin{equation}
    Y_{V}=F_{CNN}\left([\zeta_V]_{V'\in B} \right)_V.
\end{equation}

It is important to note that this is not a true probability, but a pseudo-probability calculated using the internal stream and background logit scores. For convenience, we will simply refer to this as probability from hereon. During every training epoch, voxel boxes are sampled repeatedly to reinforce the learned model. We reduce boundary effects with two separate implementations. First, by training with overlap in constructed boxes ensures regions near boundaries are sampled multiple times to produce an average reading. Secondly, we introduce a lower weighting to the edge of the boxes, to reduce their contribution to a small degree. This prevents streams or background features that have been split from producing an errant model. After training in every epoch, a validation pass is run. Here, the validation serves to remain a measure of the model's performance and has no back propagation that directly impacts the learning of the model. The model ultimately aims to reduce the Cross Entropy (CE) loss. From this, the final `best' model is selected via the MCC score.

\subsection{Pointcloud Segmentation}

In a similar vein to voxel segmentation, we can also take a patch, or pointcloud, segmentation approach. Instead of applying ML over a voxel or grid, we apply it over a sample of individual points during training. Here, instead of using a 3D convolutional network, we utilise MLP and attention to produce a learned representation for the problem. Taking inspiration from the ML methodology of PointNet++, a multilayer perceptron (MLP) is applied independently to every particle in a $k$-nearest-neighbour patch, producing a learned multi-dimensional representation for each particle. This dimensionality is a fixed architectural hyperparameter and with computational capacity taken into account, a 256-dimensional representation was chosen. As this operation uses shared weights, the representation is independent of the ordering of particles within the patch. Maximum, mean and attention-weighted pooling are then used to aggregate the particle level representations into a fixed length description of the local neighbourhood. This pooled representation is combined with the patch summary features and passed through a second MLP to produce stream and background logits for the central particle.

The initial construction of the data space is similar to the Voxel model. We produce a sampled data space containing stream and background particles. From here, the first step is to create a feature vector for the first MLP pass. To do this, we select a stream or background centre particle, $\mathbf{c}=[\mathbf{x}_{i},\mathbf{v}_i, \mathbf{L}_i]$. A kNN patch is then constructed around the chosen particle. Once again, we take a patch size of $N=48$. Next, we define all the selected neighbours in the reference frame of the centre particle. That is, the positions for particles in the patch are defined as $\mathbf{x}_j'=\mathbf{x}_j-\mathbf{x}_i$. The velocities and angular momenta are similarly computed to obtain the preliminary feature vector for the patch particles $\zeta_{j,p}=[\mathbf{x}_j',\mathbf{v}_j',\mathbf{L}_j']$. To this, we add additional patch relative features: distance to centre, $d_j'$ and speed ratio $s_j'$. At this point, all information about the patch is locally independent. To provide a more holistic context, we include the global features $[\tilde{r}_j,\tilde{|\mathbf{v}|}_j,\tilde{v}_{j,r},\tilde{v}_{j,t},\tilde{|\mathbf{L}}_j|]$. Here, each feature, $X$, is normalised across the entire training dataset as:
%$[r_j',|\mathbf{v}_j|',v_{j,r}',v_{j,t}',|\mathbf{L}_j|']$

\begin{equation}
    \label{eqn:normalisation}
    \tilde{X_i}=\frac{X_i-\mu}{\sigma},
\end{equation}

where $\mu$ and $\sigma$ are the data space's global mean and standard deviation. The motivation here is similar to that of the logic applied in the XGBoost feature creation process. We do not want the algorithm to prioritise learning absolute phase space ranges considering the dataset is synthetic and would scale differently to any independent testing data. Combining everything together gives us the patch feature vector,

\begin{equation}
    \zeta_{j,p}=[\mathbf{x}_j',\mathbf{v}_j',\mathbf{L}_j',d_j',s_j',\tilde{r}_j,\tilde{|\mathbf{v}|}_j,\tilde{v}_{j,r},\tilde{v}_{j,t},\tilde{|\mathbf{L}}_j|].
\end{equation}

From here, the first layer of MLP is applied across the patch, producing a $k\times256$ length latent vector for the $k$ particles in the patch. However, two issues arise from this. First, the high dimensionality brings excess redundancy, and second, with the iteration across many patches, the computational cost dramatically increases. As a result, we pool the created representation to reduce the dimensionality. To do so, we apply three values:
\begin{enumerate}
    \item Max Pooling -- the `strongest' activation signal: $$h_{max}=\mathrm{MAX}(h_i)$$
    \item Average Pooling -- the `average' activation signal: $$h_{mean}=\frac{1}{k}\sum{h_i}$$
    \item Attention Pooling -- weighted activation signal: $$h_{attn}=\sum\mathrm{Attention}(h_i)$$
\end{enumerate}
These three form a concatenated 3D vector: $\mathbf{h}=[h_{max},h_{mean},h_{attn}]$. This serves as a `summary' statistic of the $k$ particles in the patch, reducing the dimensionality from $k\times256\rightarrow3\times256$. To further strengthen the model and to take full advantage of the computationally costly kNN building process, we also introduce additional nearest neighbour summary features, $\mathcal{N}$ to provide an overall representation of the entire patch as opposed to individual particles in the patch. These are defined in the same fashion as was for the XGBoost feature vectors with respect to $\mathbf{c}$. We end up with the nearest neighbour vector 
\begin{equation}
    \mathcal{N}_i=[\bar{d}_i,d_{i,max},\alpha_{i,1}, \alpha_{i,2},\sigma_{i,v},\sigma_{i,v_r},s_i, \bar{v_{i,t}}, \bar{v'_{i,r}},R_i, A_i]
\end{equation} 
Along with the pooled values, this provides the summary feature vector,
\begin{equation}
    \zeta_{i,S,p}=[\mathbf{h},\mathcal{N}_i].
\end{equation}
A second MLP layer is applied over the new values and similar to the Voxel pipeline, the dimensionality is reduced to provide a stream and background logit. From this, we obtain the Pointcloud probability 
\begin{equation}
    y_{i,p}=F_{MLP}(\zeta_{i,S,p}).
\end{equation}

Similar to the Voxel model, this is also a pseudo-probability. During testing and inference, each particle in the dataset is assigned as the centre particle. As a result, each particle contributes to the model probability allocations in its local neighbourhood while also receiving its own individual value. 
To improve generalisation and decrease overfitting, random augmentations were also introduced. Given the dataset inherently covers a wide range of stream samples of varying morphology and kinematics, we did not introduce coordinate rotations or size scaling. Instead, the augmentations focused on discouraging fitting to vector patterns or over reliance on certain features. To this end, we introduced neighbour shuffling in $\zeta_{j,p}$ to ensure ordering of particles does not effect learning. Position and velocity jitters to further reduce bias to specific values and instead look for overarching representation. Furthermore, we introduced feature dropouts where the features were either set to zero or the mean to reduce excessive bias or reliance on a single feature. Finally, neural network dropout was also implemented to further discourage overfitting. The validation and model selection methodology remains identical to the Voxel model where the model reduces CE to find the model with the best MCC score. From here, we apply the model to testing.

\begin{figure}
    \centering
    \includegraphics[width=\linewidth]{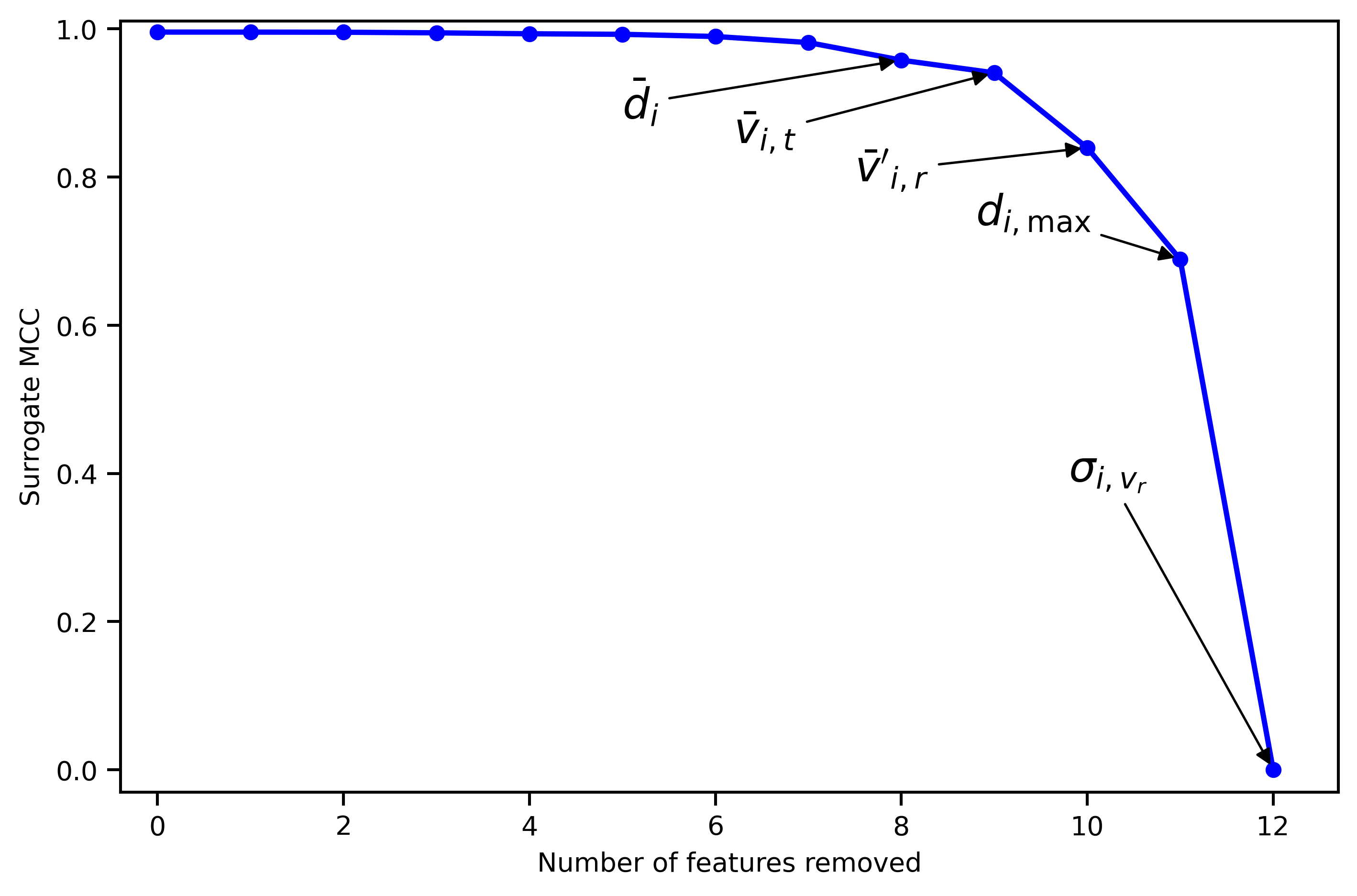}
    \caption{Ablation study conducted on the synthetic testing dataset with a surrogate model ablation for the Pointcloud algorithm. Features are incrementally removed and the performance of the model is reassessed. Therefore it is a representation of the performance drop of a particular feature given the prior removal of any other particular feature.}
    \label{fig:pc_ablation_model}
\end{figure}

Given that the Pointcloud code and XGBoost utilise nearly identical phase space derived values, we can compare the impact of each variable in a similar fashion to Fig. \ref{fig:xgboost_beeswarm}. However, since Pointcloud runs on a fundamentally different model architecture, we instead conduct a simple ablation study on our controlled synthetic testing data with the utilisation of a surrogate model. In essence, it provides a measure of performance change with the incremental removal of features. This is visualised in fig. \ref{fig:pc_ablation_model} with the five quantities resulting in performance drop as labelled. 

Interestingly, the only top five feature common between the XGBoost and Pointcloud models is the mean relative radial velocity, $\bar{v'}_{i,r}$. This is perhaps a consequence of the learned decision tree versus deep learning latent representation difference between the two codes. The XGBoost code only utilises the raw value for each feature. On the other hand, the Pointcloud model is able to utilise the given values to form its own internal functions. Thus each feature value is not necessarily important on their own, but rather contribute significantly to the internal function values. This high dimensional vector of values is what is then used to assign stream probability. We see that the radial velocity dispersion, $\sigma_{i,v_r}$ contributes the most to the internal function values. The caveat here is that this is the result of incremental feature removal and thus, this may be a breaking point in the model as it has already lost significant information via the other values. Regardless, $\bar{d}_{i},\bar{v}_{i,t},\bar{v'}_{i,r}, d_{i,max}$ and $\sigma_{i,v_r}$ can be considered performance significant values utilised in the latent functions for stream particle classification in this model. 

\begin{table*}
\centering
\caption{
Performance of the four classifiers on the halo-only and halo+structure test sets at a fixed operating threshold. Deep learning models Voxel, Pointcloud and 6D-LitePT are evaluated at pseudo-probability $y_{\rm stream}=0.5$. XGBoost uses an equivalent raw logistic probability from all 51 stored trees $y_{\rm XGB}\geq0.5$. Models are ordered by decreasing MCC.}
\label{tab:fixed_results}

\renewcommand{\arraystretch}{1.25}
\setlength{\tabcolsep}{8pt}

\begin{tabular*}{\textwidth}{@{\extracolsep{\fill}}llcccccc}
\toprule
Test Set & Model & Threshold & Precision & Recall & F1 & IoU & MCC \\
\midrule

\multirow{4}{*}{Halo Only}
& Pointcloud & 0.50 & 0.988 & 0.975 & 0.982 & 0.964 & \textbf{0.980} \\
& XGBoost    & 0.50 & 0.988 & 0.893 & 0.938 & 0.883 & \textbf{0.934} \\
& 6D-LitePT  & 0.50 & 0.978 & 0.832 & 0.899 & 0.816 & \textbf{0.898} \\
& Voxel      & 0.50 & 0.788 & 0.657 & 0.717 & 0.558 & \textbf{0.717} \\
\addlinespace[0.75ex]
\midrule

\multirow{4}{*}{Halo + Structures}
& 6D-LitePT  & 0.50 & 0.844 & 0.815 & 0.829 & 0.708 & \textbf{0.825} \\
& Voxel      & 0.50 & 0.590 & 0.505 & 0.544 & 0.374 & \textbf{0.542} \\
& Pointcloud & 0.50 & 0.193 & 0.977 & 0.322 & 0.192 & \textbf{0.328} \\
& XGBoost    & 0.50 & 0.197 & 0.898 & 0.324 & 0.193 & \textbf{0.311} \\
\bottomrule
\end{tabular*}
\end{table*}

\begin{table*}
\centering
\caption{Performance of the four classifiers on the halo-only and halo+structure test sets. For Voxel, Pointcloud and 6D-LitePT, thresholds correspond to the operating point that maximises the validation MCC. XGBoost retains the use of a raw logistic probability threshold for stored trees. Models are ordered by decreasing MCC.}
\label{tab:best_results}

\renewcommand{\arraystretch}{1.25}
\setlength{\tabcolsep}{8pt}

\begin{tabular*}{\textwidth}{@{\extracolsep{\fill}}llcccccc}
\toprule
Test Set & Model & Threshold & Precision & Recall & F1 & IoU & MCC \\
\midrule

\multirow{4}{*}{Halo Only}
& Pointcloud & 0.60 & 0.992 & 0.973 & 0.983 & 0.966 & \textbf{0.981} \\
& XGBoost    & 0.50 & 0.988 & 0.893 & 0.938 & 0.883 & \textbf{0.934} \\
& 6D-LitePT  & 0.30 & 0.960 & 0.852 & 0.903 & 0.823 & \textbf{0.901} \\
& Voxel      & 0.60 & 0.835 & 0.620 & 0.712 & 0.552 & \textbf{0.717} \\
\addlinespace[0.75ex]
\midrule

\multirow{4}{*}{Halo + Structures}
& 6D-LitePT  & 0.50 & 0.844 & 0.815 & 0.829 & 0.708 & \textbf{0.825} \\
& Voxel      & 0.30 & 0.527 & 0.575 & 0.550 & 0.379 & \textbf{0.546} \\
& Pointcloud & 0.95 & 0.205 & 0.941 & 0.337 & 0.202 & \textbf{0.336} \\
& XGBoost    & 0.50 & 0.197 & 0.898 & 0.324 & 0.193 & \textbf{0.311} \\
\bottomrule
\end{tabular*}
\end{table*}

\subsection{6D-LitePT: Hybrid Classifier}

The initial configuration remains the same as we construct a data space containing stream and background particles. Next, we select a sample of centre particles and construct a box around them. From here, we define the particle features in a similar fashion to our Pointcloud approach. Each particle has their phase space enriched to $[\mathbf{x}_i,\mathbf{v}_i,r,v_{i,r},v_{i,t},\mathbf{L}_i,|L_i|]$. From here, they undergo normalisation, with spatial coordinates defined as $\mathbf{x_i}'=\mathbf{x}_i-\mathbf{x}_c$ and all other features modified as per Eqn. \ref{eqn:normalisation}. Next, the box is voxelised. However, unlike the Voxel segmentation model, here, we voxelise in 6D phase space as $[\mathbf{x}_i,w_v\mathbf{v}_i]$, where $w_v$ is a weighting on how much the velocity contributed. We use an empirical value of $w_v=0.05$ here, which we find to be effective for the given model. This voxelisation separates particles jointly by position and velocity, allowing spatially close populations with different velocities to occupy distinct cells. Within each voxel, we once again average the values providing the cell level feature vector,

\begin{equation}
    \zeta_{V,LPT}=[\bar{\mathbf{x}'},\bar{\tilde{\mathbf{v}}},\bar{\tilde{r}},\bar{\tilde{v_{r}}},\bar{\tilde{v_{t}}},\bar{\tilde{\mathbf{L}}},\bar{\tilde{|L|}}].
\end{equation}

Given the introduction of velocity space to the voxelisation process, we are able to collect similar kinematic features to the XGBoost and Pointcloud frameworks. Thus for the voxel, we introduce the summary features: log particle count, velocity dispersion, radial velocity dispersion, speed dispersion, angular momentum alignment, angular momentum magnitude dispersion giving us the nearest neighbour vector for each voxel:
\begin{equation}
    \mathcal{N}_V =\left[\log N_V,\sigma_{V,v},\sigma_{V,v_r},\sigma_{V,|\mathbf{v}|},A_{V},\sigma_{V,|L|}\right].
\end{equation}
Combined with the particle level values, gives us the LitePT feature vector:
\begin{equation}
    \zeta_{LPT}=[\zeta_{V,LPT},\mathcal{N}_V].
\end{equation}
\begin{figure*}
    \centering
    \setlength{\tabcolsep}{1pt}
    \renewcommand{\arraystretch}{0}

    \begin{tabular}{@{}ccc@{}}
        \includegraphics[width=0.328\textwidth]
            {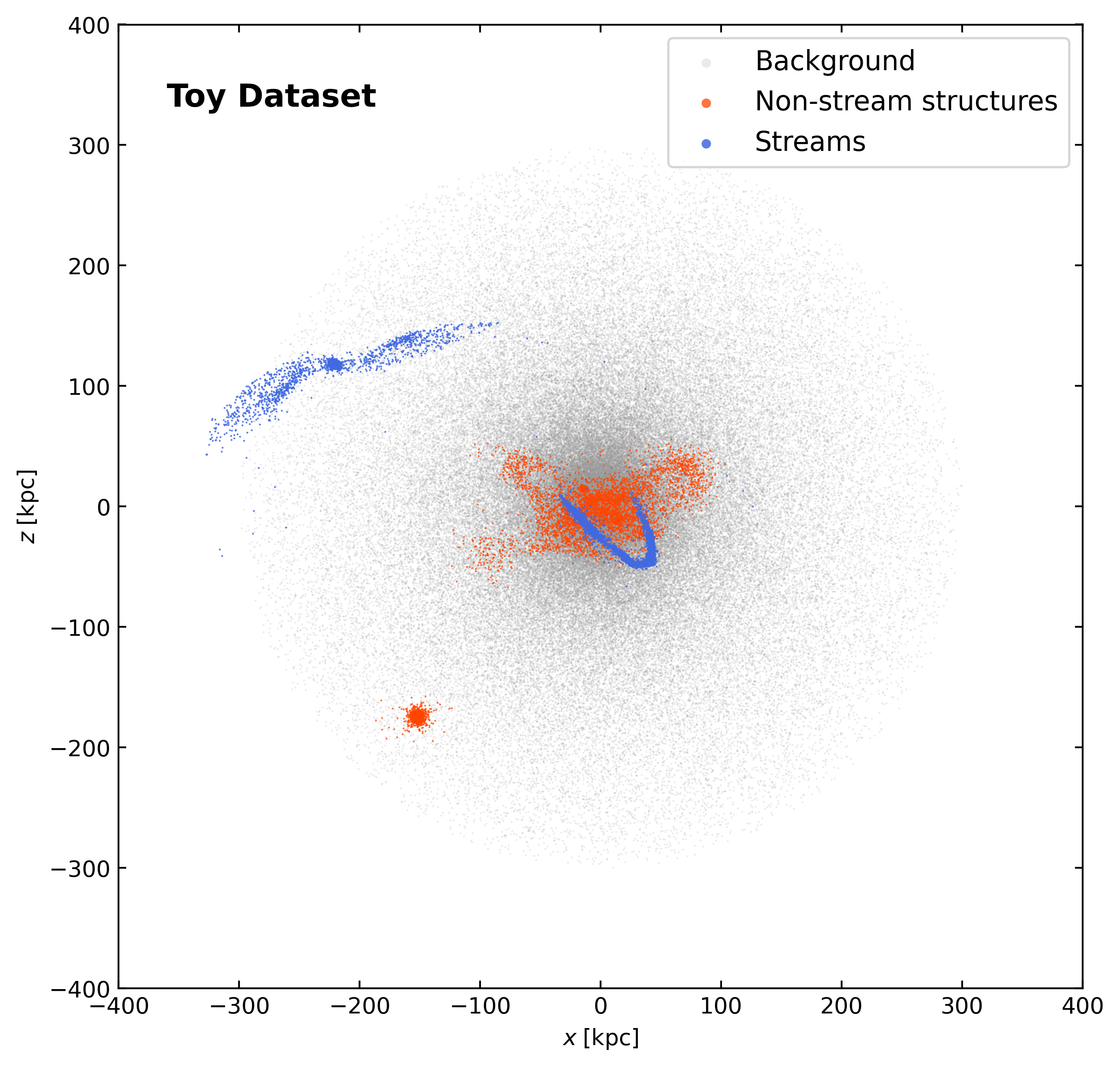} &
        \includegraphics[width=0.328\textwidth]
            {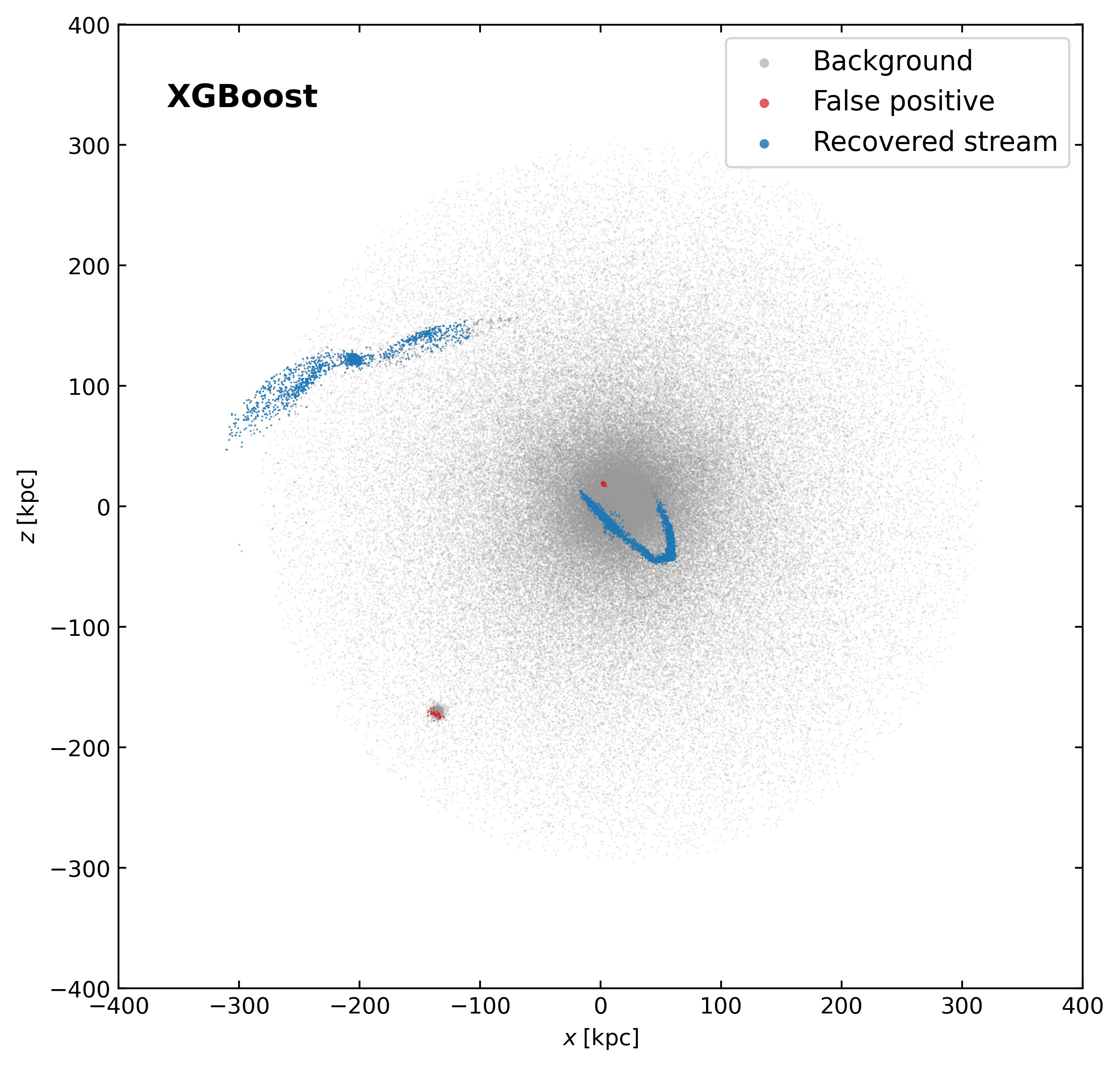} &
        \includegraphics[width=0.328\textwidth]
            {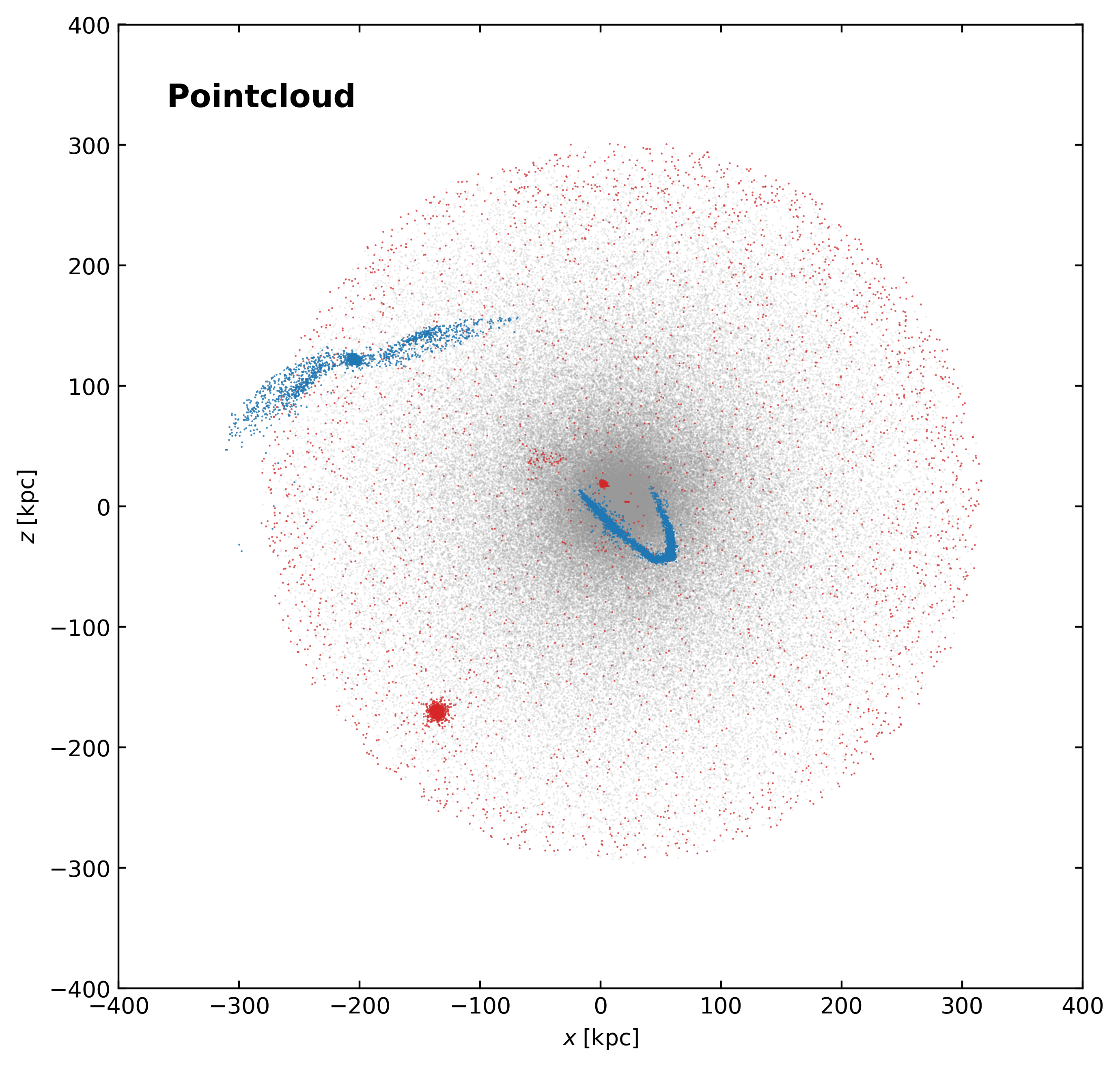}
        \\[-2pt]
        \includegraphics[width=0.328\textwidth]
            {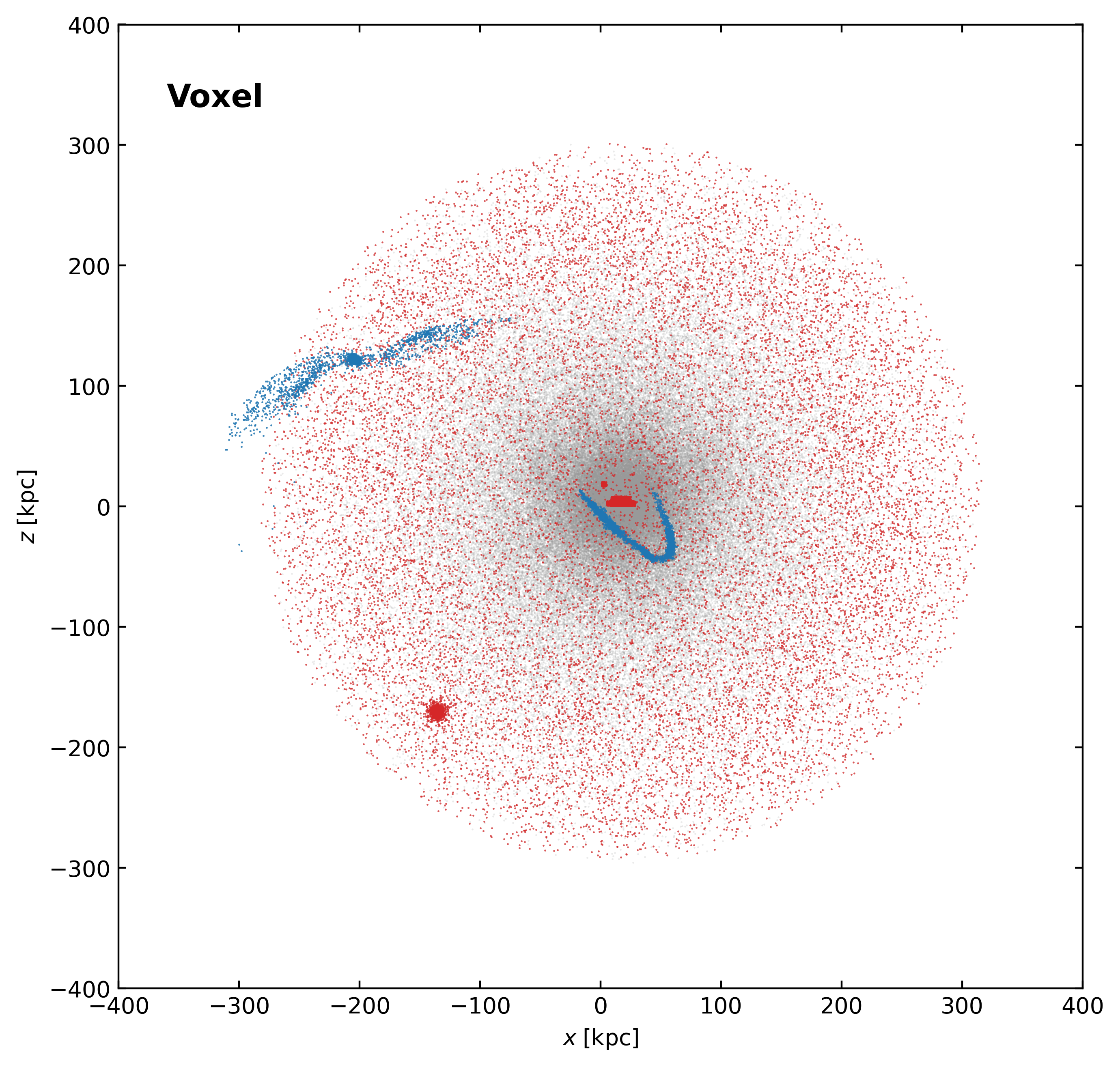} &
        \includegraphics[width=0.328\textwidth]
            {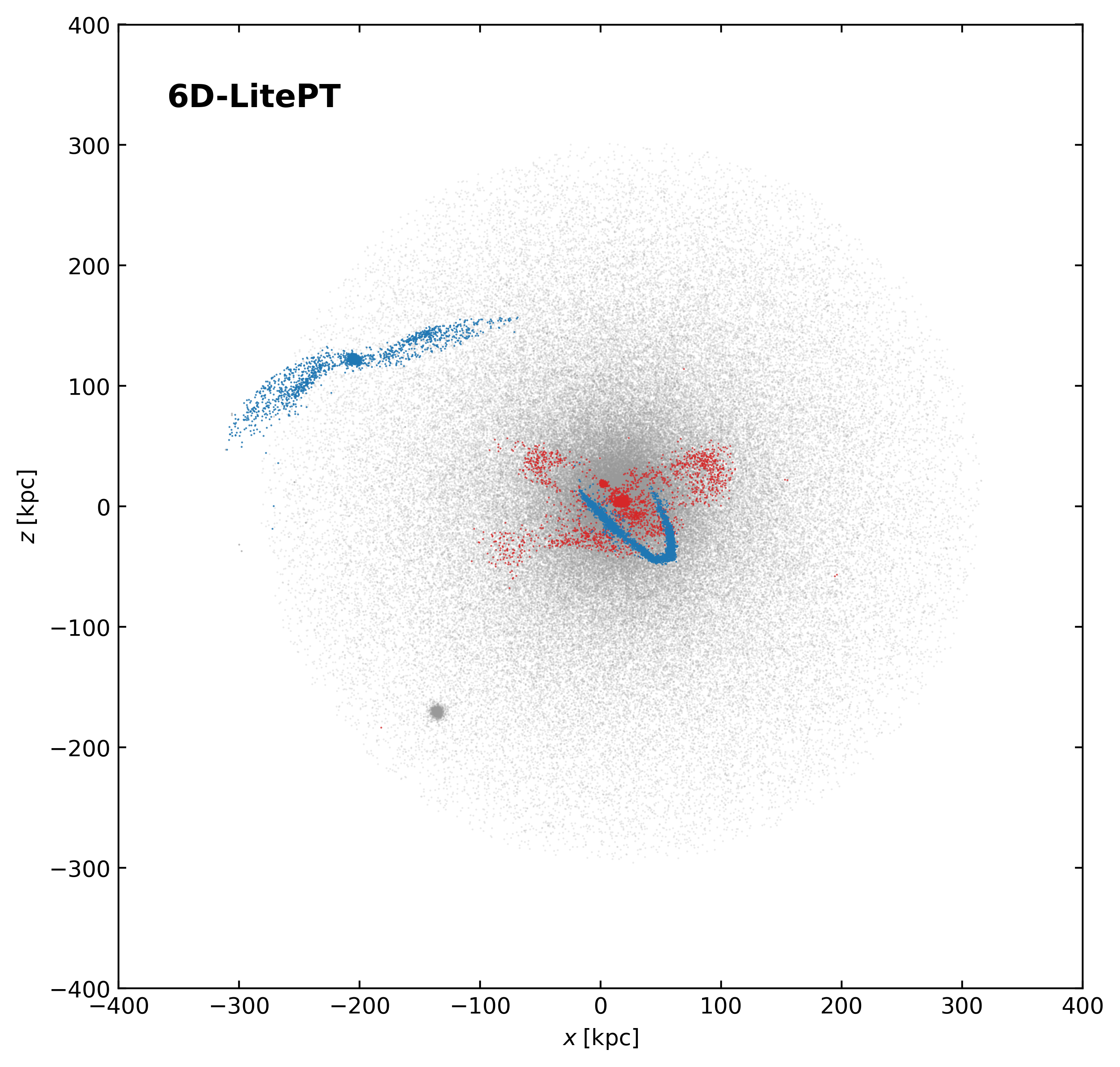} &
        \includegraphics[width=0.328\textwidth]
            {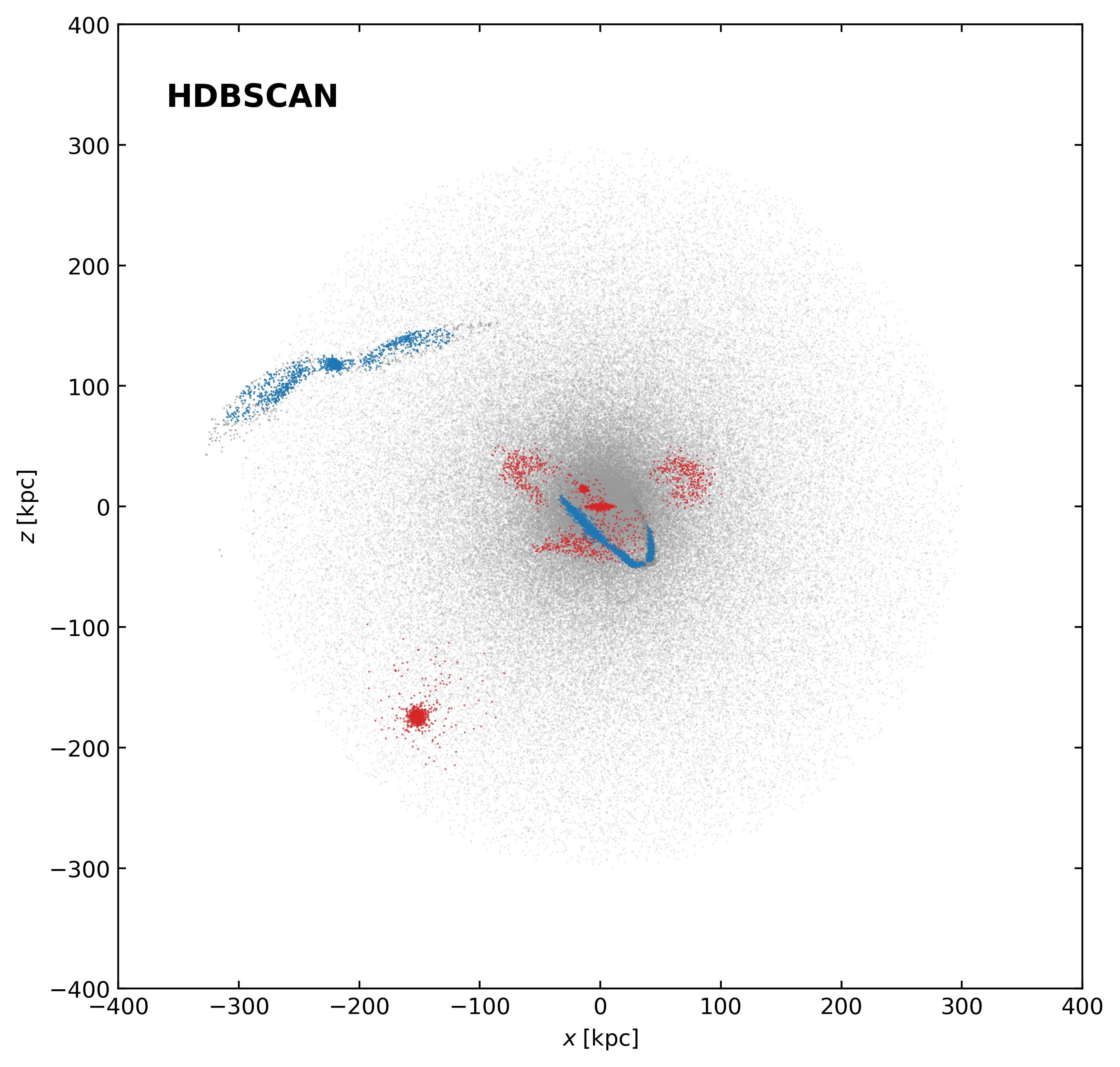}
    \end{tabular}

    \caption{Comparison of stream-particle recovery in the toy dataset using
    the four presented machine-learning models and the HDBSCAN clustering
    algorithm.}
    \label{fig:toy_stream_recovery}
\end{figure*}
Here, we deviate from both the Pointcloud and Voxel pipelines through the use of the LitePT transformer. Notably, there is no application of a CNN. Instead, the model applies serialised local attention to sparse tokens representing occupied cells in six-dimensional phase space. Its attention neighbourhoods are therefore defined jointly by particle positions and velocities, rather than by spatial proximity alone with velocity supplied only as an auxiliary feature. MLP layers learn relationships between neighbouring phase space tokens, while hierarchical pooling progressively aggregates information over larger regions. Decoder unpooling then restores the token resolution, and a linear segmentation head produces stream and background logits. In addition to its use of attention, LitePT offers the advantage of computational efficiency. Its sparse coordinate representation stores and processes only occupied phase space cells, thereby reducing the memory overhead that would be incurred by a dense 6D grid. Finally, although the computation is performed over voxel tokens, the resulting probabilities are inverse mapped to the particles contained within each cell to provide particle specific predictions. This gives the LitePT hybrid model probability
\begin{equation}
    y_{i,LPT}=F_{LitePT}\left([\zeta_{LPT}]_{V'\in B} \right)_i.
\end{equation}
As was for Voxel and Pointcloud, this is also a pseudo-probability. One final point of difference in the LitePT model is the use of Fractal Tanimoto with Negation and Multiple Topologies \citep[FTNMT;][]{Foivos_Tanimoto_loss_2021,Chatterjee_2021} loss as opposed to CE as we see better performance with its implementation. However, MCC is still utilised as the model selection criteria. 

\subsection{Testing and Comparison}

Once trained and validated, the models were applied to a testing dataset. Inference was conducted over the complete dataset to ensure each particle receives a model probability score. For this purpose, the XGBoost and Pointcloud codes assigned each particle as a centre particle for computation of their individual probabilities. The Voxel and 6D-LitePT models utilised systematic overlapping grid sampling across the entire dataset to increase resistance against boundary effects. The results of testing can be seen in tables \ref{tab:fixed_results} and \ref{tab:best_results}. Table \ref{tab:fixed_results} presents a fixed threshold to compare the models strictly against each other while table \ref{tab:best_results} presents the validation selected thresholds for the deep learning models, with XGBoost retaining the raw logistic probability threshold of $0.5$.

The MCC value shows a strong result for the XGBoost model in the halo only case, with a score of $0.934$ compared with $0.898$ for 6D-LitePT at the fixed threshold. Importantly, with the introduction of other non-stream structures, we observe a decrease of $0.623$ in testing MCC to $0.311$, indicating substantial sensitivity to added structure in this particular test. F1 and Intersection over Union (IoU) reflect a similar change. In contrast we see that the Voxel model performs the poorest in the halo only scenario, but achieves the second highest MCC in the halo + structure scenario, with fixed threshold scores of $0.717$ and $0.542$, respectively. One possible limitation is the compression of each spatial cell into particle count and mean kinematic quantities. When stream and background particles occupy the same cell, these averages can dilute the stream signature, while a shared voxel prediction cannot distinguish the populations within it. Compounding the issue, the use of averaged values over nearest neighbour informed particle features gives the voxel model inherently less stream specific dynamical phase space information. As will be discussed later, we address this in the 6D-LitePT model which similarly utilises voxels.

The Pointcloud model also displays drastically different performances under the two testing conditions. At best threshold, in the halo only case, it performs almost perfectly, with an exceptional score of $0.981$ at a threshold of $0.6$. However, the introduction of additional structures drastically reduces MCC to $0.336$ at the validation selected threshold, below 6D-LitePT and Voxel but slightly above XGBoost. This provides an insight into the model selection process. We see that it is able to detect kinematic coherence from background extremely well but struggles to separate structures with more nuanced phase space distribution changes. Once again, it is important to note that the models receive no temporal context. Thus, from the perspective of the model, structures such as shells in their intermediary phase may share spatial and kinematic features with a stream in a particularly eccentric orbit. Furthermore, we suspect the model misinterprets slightly disturbed subhalos that have yet to be meaningfully disrupted as the central cores of a stellar stream. This is a particularly ambiguous scenario as these subhalos may eventually produce tails and convert into a stellar stream, had they had time to evolve. Despite the susceptibility to background contamination, it is worth noting that in both scenarios, it is the strongest model at recovering stream particles with a recall (stream recovery fraction) of $0.975$ and $0.977$ in the fixed threshold regime.

Finally, inspecting the results of the 6D-LitePT model, we see that it has the most robust performance during this particular test. It is able to effectively function in both halo and background structure contaminated scenarios. At $0.073$, it has the lowest MCC score drop in the fixed threshold measure. Its performance in the halo + structure case is supported by its stream detection purity. In other words, of all particles identified as stream-like, it has the highest precision in this scenario, at $0.844$. This indicates that the 6D-LitePT model produces the least contaminated stream selection in the halo + structure test. However, its major limitation in this test is detecting stream-like particles, with the third highest recall, at $0.815$, trailing behind the XGBoost and Pointcloud models.

\begin{table}
\centering
\caption{Precision, recall and MCC of ML models and HDBSCAN on the toy dataset. Since HDBSCAN is a general structure finder, it has no stream finding precision. Instead, its purity is simply the measure of how much of the identified structure were streams. The highest value in each column is highlighted in bold.}
\label{tab:toy_recovery}

\renewcommand{\arraystretch}{1.25}
\setlength{\tabcolsep}{6pt}

\begin{tabular*}{0.95\columnwidth}{@{\extracolsep{\fill}}lccc@{}}
\toprule
Model & Precision & Recall & MCC \\
\midrule
XGBoost    & \textbf{0.901} & 0.968          & \textbf{0.927} \\
Pointcloud & 0.579          & \textbf{0.998} & 0.732 \\
Voxel      & 0.269          & 0.970          & 0.433 \\
6D-LitePT  & 0.633          & 0.997          & 0.771 \\
HDBSCAN    & 0.502\rlap{*}  & 0.934          & 0.645 \\
\bottomrule
\end{tabular*}
\end{table}

To visualise these results, we conduct inference on a simple toy dataset and compare the performance of the models with a separate stream detection methodology. For this purpose, we utilise HDBSCAN \citep{McInnes2017}, which, while not a specialised stream detection tool, it is an effective density clustering algorithm capable of separating particle structure from background noise. Fig. \ref{fig:toy_stream_recovery} provides a visualisation of the toy dataset and model performances, including HDBSCAN, which receives the full six-dimensional phase space coordinates. Hence it is provided the same underlying positional and kinematic information available to the ML models. Each coordinate is standardised to zero mean and unit variance before clustering with a Euclidean metric, a minimum cluster size of 200, a minimum samples parameter of 20, and leaf cluster selection. We include two streams, close and at a distance from the centre as well as a coherent disrupted structure and an undisturbed subhalo core. Table \ref{tab:toy_recovery} details the precision - the purity of the selected stream sample, and recall - the fraction of selected true stream particles. A higher precision entails less background contamination and a higher recall equates to greater stream recovery.

XGBoost presents itself as the best performer, with a recall of $0.968$ and precision of $0.901$. The Pointcloud model retains its place as the strongest stream recovery tool with a recall of $0.998$. We see that it struggles to filter out the undisturbed halo core, contributing the bulk of the contamination. In contrast, the 6D-LitePT model successfully filters out the complete disturbed core, though struggling with the non-stream structure. The voxel model is contaminated by the toy background halo as well as the undisturbed core and has the greatest contamination out of the five models presented. HDBSCAN, being a general clustering algorithm, is not able to distinctly identify stream structure. Instead, we manually separated clusters belonging to the streams and other identified structure with the known ground truth stream particles to determine the recall and precision. As a result, purity is not as relevant for the algorithm. However, its recovered stream percentage still lags behind the ML models. Considering also the manual input for structure separation, as a tool for stream classification, we see that all four ML algorithms exceed HDBSCAN in recall performance. All testing thus far have been conducted on artificially produced datasets with no temporal or dynamical correlation. For the inspection of a more practical scenario, we next apply the four ML models to the Auriga suite, in particular Au-24, as well as a summarised inspection of Au-6,16,21,23,24,27.

\section{Independent Dataset Testing}
%\textcolor{red}{\section{Independent Dataset Testing}}
\label{sec:independent_dataset_testing}
\subsection{Au-24 Inference}

We evaluate the models trained on the synthetic dataset using an independent cosmological simulation from the Auriga suite labelled catalogue \citep{Riley_2025}. Here, we present Au-24 at level-3 resolution at $z=0$. This was chosen as an example for the visual clarity of the present stream population. Averaging MCC across all four models, the strongest performer was Au-23 and the weakest was Au-6. Both of these results are provided in Appendix \ref{appendix} along with Au-16,21 and 27. These simulations have a baryonic mass resolution of $6.7\times10^3\,{\rm M_\odot}$, and a dark matter particle mass of $3.6\times10^4\,{\rm M_\odot}$. This application tests whether the spatial and kinematic representations learned from the synthetic structures transfer to a cosmological environment. As such, here, the original trained models are applied without retraining on Auriga.

The Auriga catalogue provides reference classifications for the accreted systems, separating them into intact satellites, streams and phase-mixed structures. We assign each particle the classification of its parent system, with particles belonging to stream classified systems treated as positive examples and intact, phase-mixed and in-situ particles treated as negative examples. These reference classifications apply to entire systems, including any surviving bound progenitor and associated debris. As a result, an important caveat must be addressed. The metrics presented here measure the recovery of particles belonging to stream classified systems, rather than exclusively the recovery of unbound tidal tails or other stream like features. As acknowledged by \cite{Riley_2025}, the boundaries between the classified structures are not clear cut with most accreted satellites undergoing some form of disruption. Furthermore, the  Auriga classifications also differ from the morphological selection used to construct our synthetic training dataset. Systems with visible tidal features may qualify as streams in our training sample while remaining classified as intact under the Auriga bound fraction criterion. However, for consistency, and for a defined quantitative analysis, we treat the identified stream structures as the only positive ground truth. Hence, the reported metrics assess agreement with the Auriga reference classifications, with this difference in labelling potentially contributing to some disagreements.

\begin{figure*}
    \centering
    \includegraphics[width=\linewidth]{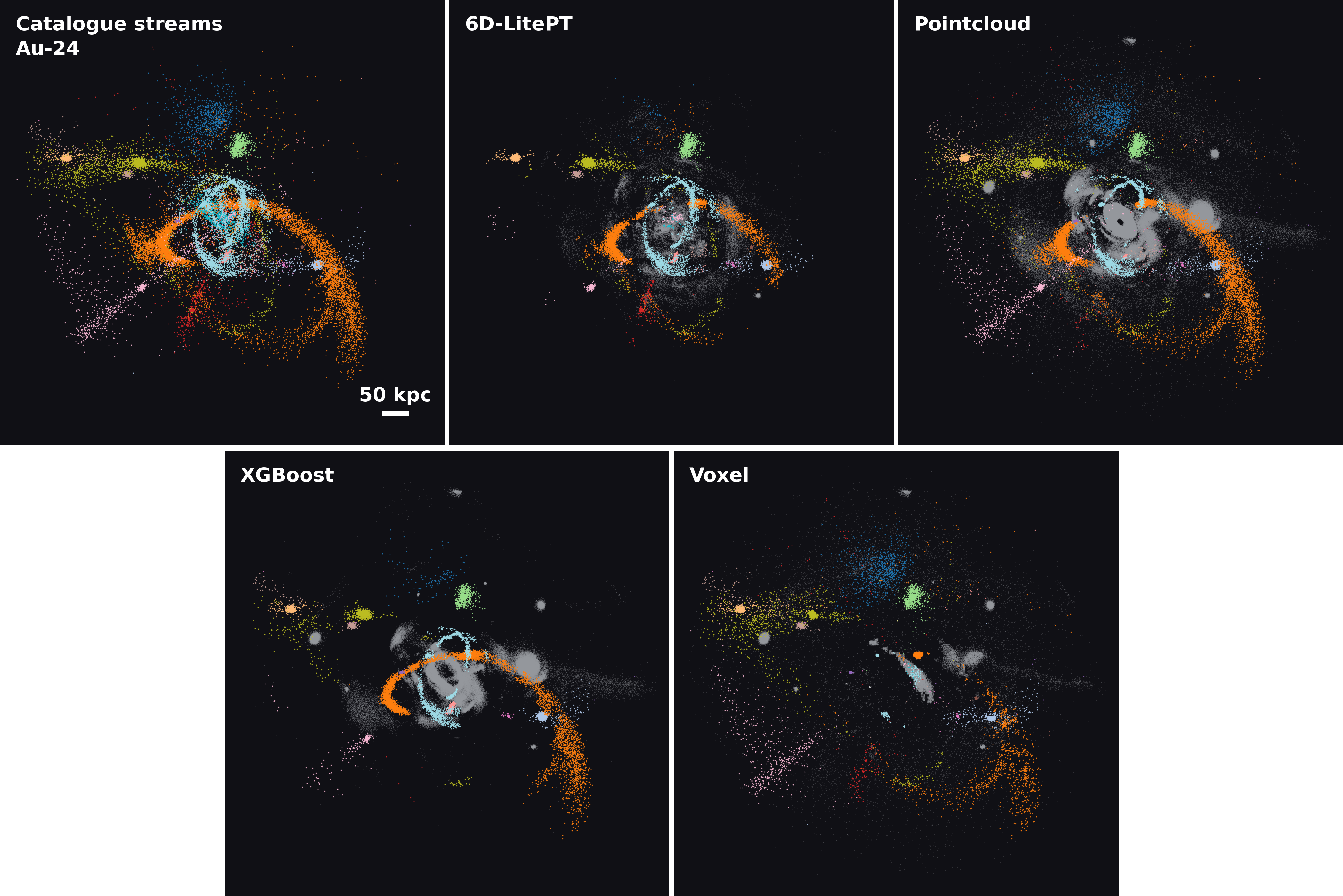}
    \caption{Comparison of the catalogue stream population and the particles selected by the original synthetic trained models in Au-24 at level-3 resolution at $z=0$, projected onto the $y$--$z$ plane. The first panel shows particles belonging to catalogue classified stream systems, with distinct colours identifying individual systems. The remaining panels show particles classified as streams by each model. Coloured particles belong to the reference stream systems, using the same colours as the first panel, while grey particles belong to non-stream reference populations.}
    \label{fig:auriga_stream_comparison}
\end{figure*}

Fig. \ref{fig:auriga_stream_comparison} compares the particle distributions selected by the four models with the reference stream population. Table \ref{tab:auriga24_recovery} quantifies their precision and recall. The 6D-LitePT model achieves the strongest balance between recovery and contamination, with a precision of $0.550$ and recall of $0.744$. Pointcloud achieves the highest recall, recovering $0.838$ of the reference stream particles, but its precision of $0.015$ indicates substantial contamination. XGBoost achieves a precision of $0.023$ and recall of $0.754$, while the Voxel model achieves a precision of $0.063$ and recall of $0.563$. Thus, although all four models recover substantial fractions of the reference stream population, their ability to reject non-stream particles differs considerably.

Table \ref{tab:auriga24_category_selection} separates this behaviour by reference category. Each entry measures the fraction of particles selected within that category, allowing stream recovery to be considered alongside the selection of intact, phase-mixed and in-situ populations. Here, we obtain an insight into the structure specific contamination encountered by all four models. Excluding 6D-LitePT, the other three models receive the majority of their contamination percentage wise from intact structures. The Pointcloud model performs relatively worse in both other non-stream categories compared to the other three models with approximately $7\%$ and $3.5\%$ contamination from in-situ and phase-mixed respectively. We however, do see strong stream identification from all four models, particularly Pointcloud with the highest stream recovery of $83.75\%$, demonstrating transfer of learned stream behaviour from the synthetic dataset. The majority selection of particles from intact systems represents a limitation under the adopted reference classification. Differences from the morphological training labels may contribute to disagreements for systems with weak tidal features, but do not establish that these selections are correct. This application gives an insight into both the recovery achievable without specialised cosmological retraining and the limitations that remain when applying the presented models to an independent simulation.

\begin{figure*}
    \centering
    \includegraphics[width=\linewidth]{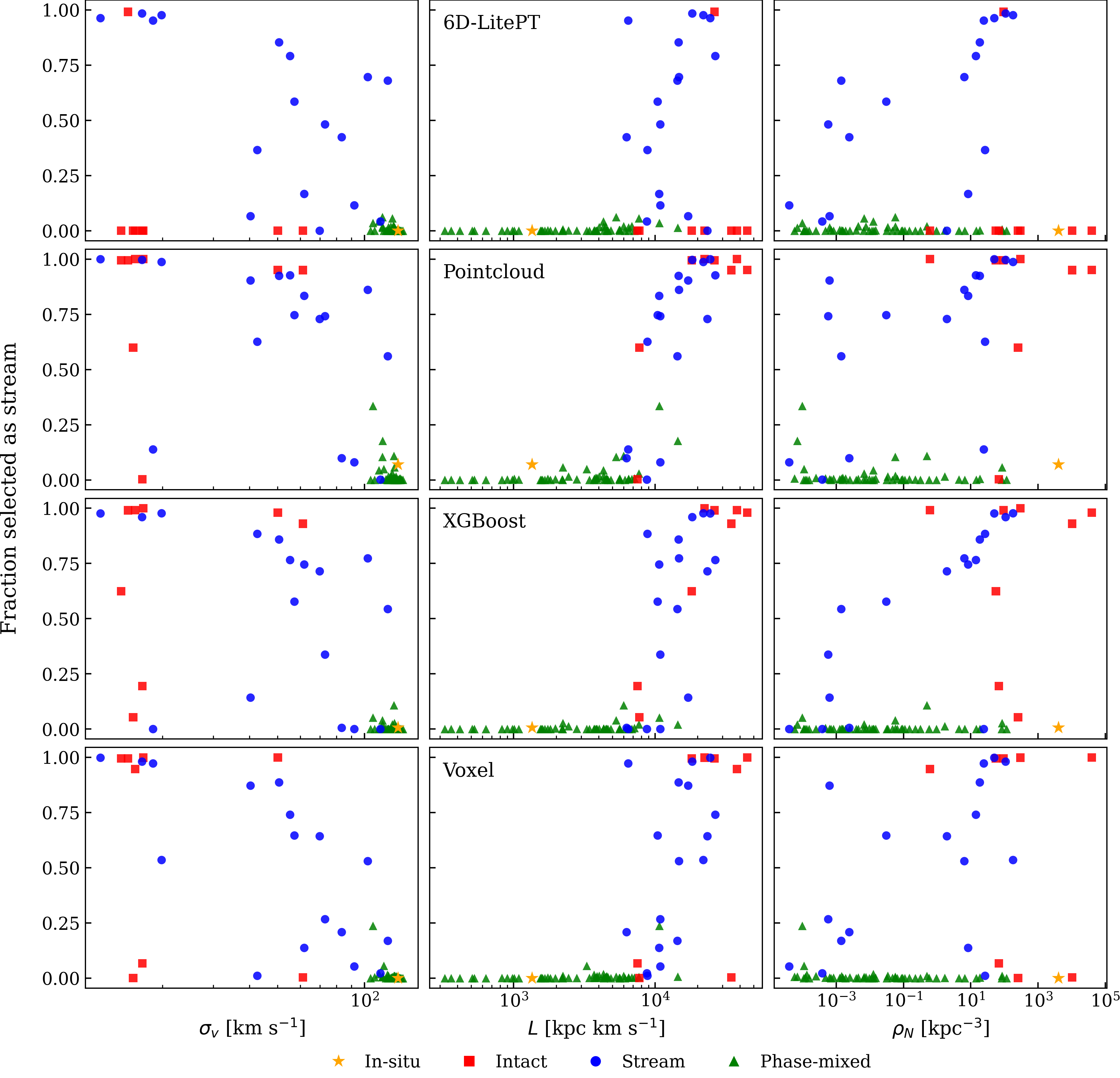}
    \caption{Fraction of evaluated particles selected as streams in Au-24 versus velocity dispersion $\sigma_v$, specific angular momentum $L$, and particle number density $\rho_N$. Rows correspond to the four original synthetic trained models. Each marker represents an accreted system: streams (blue circles), intact satellites (red squares), and phase-mixed systems (green triangles). The in-situ population is represented by one aggregate point (orange star) in each panel. The selected fraction is calculated relative to the evaluated particle count of each system or aggregate population.}
    \label{fig:auriga_system_properties}
\end{figure*}

\begin{table}
\centering
\caption{ Precision, recall and MCC of the original ML models applied to Au-24 at level-3 resolution. The highest value in each column is highlighted in bold.}
\label{tab:auriga24_recovery}

\renewcommand{\arraystretch}{1.25}
\setlength{\tabcolsep}{6pt}

\begin{tabular*}{0.95\columnwidth}{@{\extracolsep{\fill}}lccc@{}}
\toprule
Model & Precision & Recall & MCC \\
\midrule
XGBoost    & 0.023          & 0.754          & 0.123 \\
Pointcloud & 0.015          & \textbf{0.838} & 0.101 \\
Voxel      & 0.063          & 0.563          & 0.184 \\
6D-LitePT  & \textbf{0.550} & 0.744          & \textbf{0.639} \\
\bottomrule
\end{tabular*}

\end{table}

\begin{table}
\centering
\caption{
Percentage of particles selected as streams within each reference category
in Au-24 at level-3 resolution, using the original models.
Each percentage is calculated relative to the evaluated particle population
of that category. The stream column corresponds to recall, while the remaining
columns give category specific false positive rates.
The highest stream recall is highlighted in bold.
}
\label{tab:auriga24_category_selection}

\begingroup
\small
\renewcommand{\arraystretch}{1.2}
\setlength{\tabcolsep}{2pt}

\begin{tabular*}{\columnwidth}{@{\extracolsep{\fill}}lcccc@{}}
\toprule
Model & In-situ (\%) & Intact (\%) & Stream (\%) & \shortstack{Phase-\\mixed} (\%) \\
\midrule
6D-LitePT  & 0.022 & 0.119  & 74.347          & 0.857 \\
Pointcloud & 6.966 & 94.799 & \textbf{83.750} & 3.496 \\
XGBoost    & 0.644 & 93.732 & 75.438          & 1.864 \\
Voxel      & 0.025 & 25.797 & 56.310          & 0.619 \\
\bottomrule
\end{tabular*}

\endgroup
\end{table}

Taking a closer look at Fig. \ref{fig:auriga_stream_comparison}, we see the different recovery and contamination patterns of the four models. The 6D-LitePT selection most closely isolates the reference stream population, consistent with its higher precision in Table \ref{tab:auriga24_recovery}. It retains substantial portions of the prominent stream structures while rejecting many of the compact non-stream systems selected by the other models. Pointcloud recovers more extended and diffuse portions of several streams, consistent with its highest particle level recall. However, this broader selection also includes numerous intact satellites and substantial contamination near the host centre. XGBoost achieves the second lowest stream precision, retaining compact components and sections of the prominent outer structure while missing much of the more dispersed debris. Its false positives include several compact satellites and a distinct ring-like feature in projection near the host centre. The Voxel model also recovers substantial stream structure but ranks the lowest in recall amongst the four models. Its precision is marginally superior to the XGBoost and Pointcloud models. Similar to those models, Voxel selects numerous non-stream particles, including intact satellites and a portion of the central population. 

To examine these behaviours further, Fig. \ref{fig:auriga_system_properties} relates the fraction of particles selected within each system to its velocity dispersion, specific angular momentum and particle number density. Here, $\sigma_v=[\langle|\mathbf{v}-\langle\mathbf{v}\rangle|^2\rangle/3]^{1/2}$ is the 1D equivalent velocity dispersion. $L=\mathrm{median}(|\mathbf{r}\times\mathbf{v}|)$ is the median specific angular-momentum magnitude, measured in the host-centred frame. $\rho_N=N(<r_{50})/(4\pi r_{50}^3/3)$ is the mean particle number density within the half-number radius $r_{50}$, measured about the component wise median position. 

Stream and intact systems overlap substantially in these properties, particularly at low system level velocity dispersion. Pointcloud, XGBoost and Voxel frequently assign high selected fractions to both populations in this regime. This behaviour is consistent with sensitivity to structures with relatively small internal velocity spreads, a property shared by streams and intact satellites. In contrast, 6D-LitePT rejects most intact systems while retaining high selected fractions for several streams occupying similar ranges of the plotted properties. Its discrimination therefore cannot be described by a simple threshold in velocity dispersion alone. Within the stream population, larger selected fractions generally occur towards higher specific angular momentum and number density. Several low density streams are only partially recovered or largely missed, whereas more concentrated systems are often selected more completely. Concentration alone, however, does not distinguish streams from intact satellites. This is highlighted by the 6D-LitePT model as it rejects several dense intact systems unlike the other three models. Similarly, the association with angular momentum may partly reflect differences between the reference populations rather than an independent dependence of selection on angular momentum. 

Most phase-mixed systems have small selected fractions for all four models. Collectively, the diagnostics suggest that low system level velocity dispersion and concentration are associated with high selected fractions for several models, while 6D-LitePT achieves greater separation between streams and intact systems with overlapping bulk properties. It is important to note that these associations describe the observed predictions but do not uniquely identify the internal classification rules learned by the models. Overall, the models recover stream structure while rejecting most in-situ and phase-mixed particles. This indicates that the models are able to discern stream like structure from background and other coherent non-stream and non-intact structure. However, confusion with intact satellites as defined for the Auriga suite remains a prominent limitation, particularly for Pointcloud and XGBoost.

\subsection{Auriga Suite Inference}
\label{subsec:auriga_suite_inference}
\begin{figure}
    \centering
    \includegraphics[width=\linewidth]{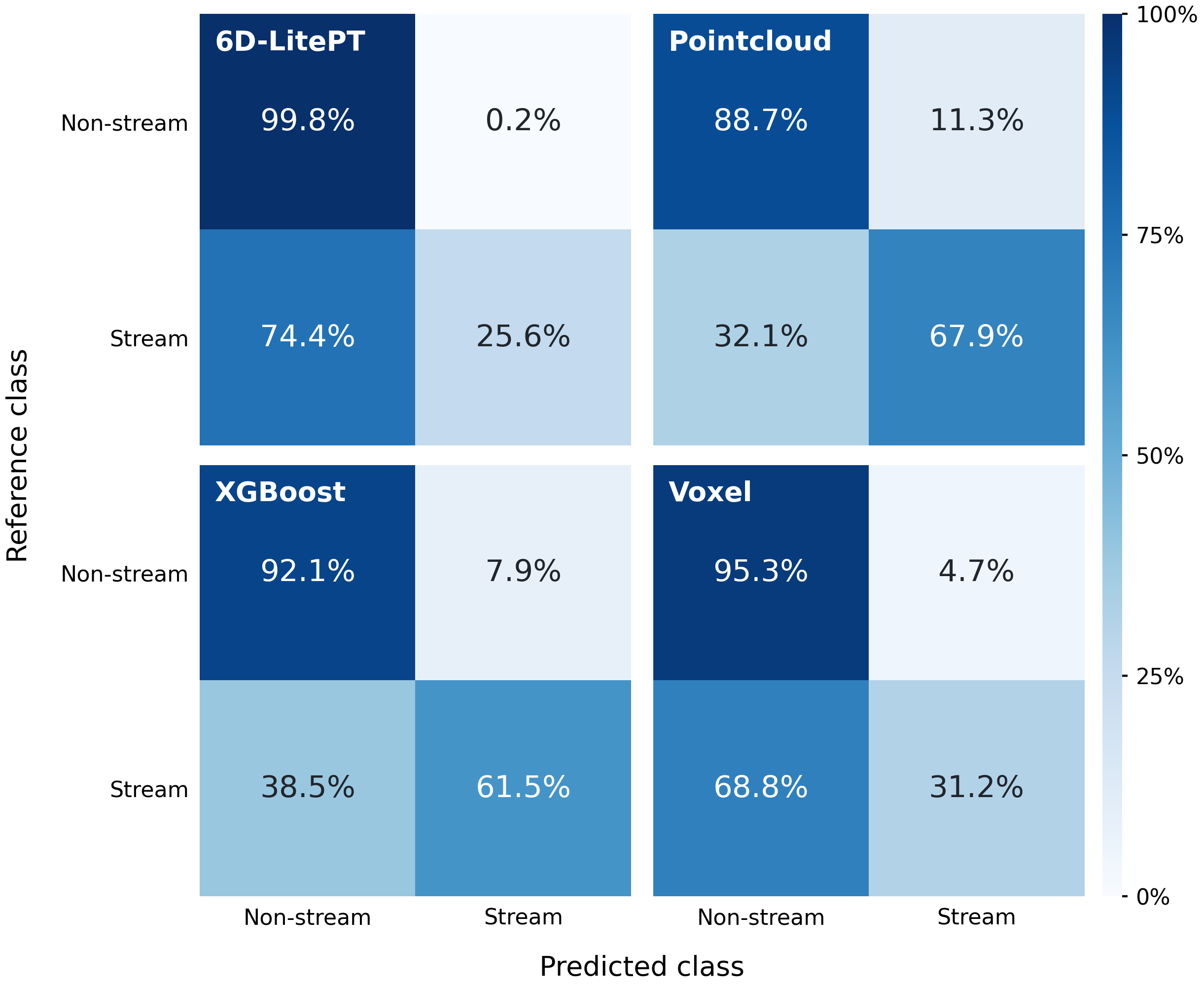}
    \caption{Confusion matrices for the four models evaluated on the six halos in the Auriga suite. Catalogue stream system members constitute the positive `Stream' class while intact, phase-mixed and in-situ particles constitute the `Non-stream' class. Percentages are calculated within each halo and then averaged with equal weight across the six halos.}
    \label{fig:confusion_matrices}
\end{figure}

\begin{figure*}
    \centering
    \includegraphics[width=\linewidth]{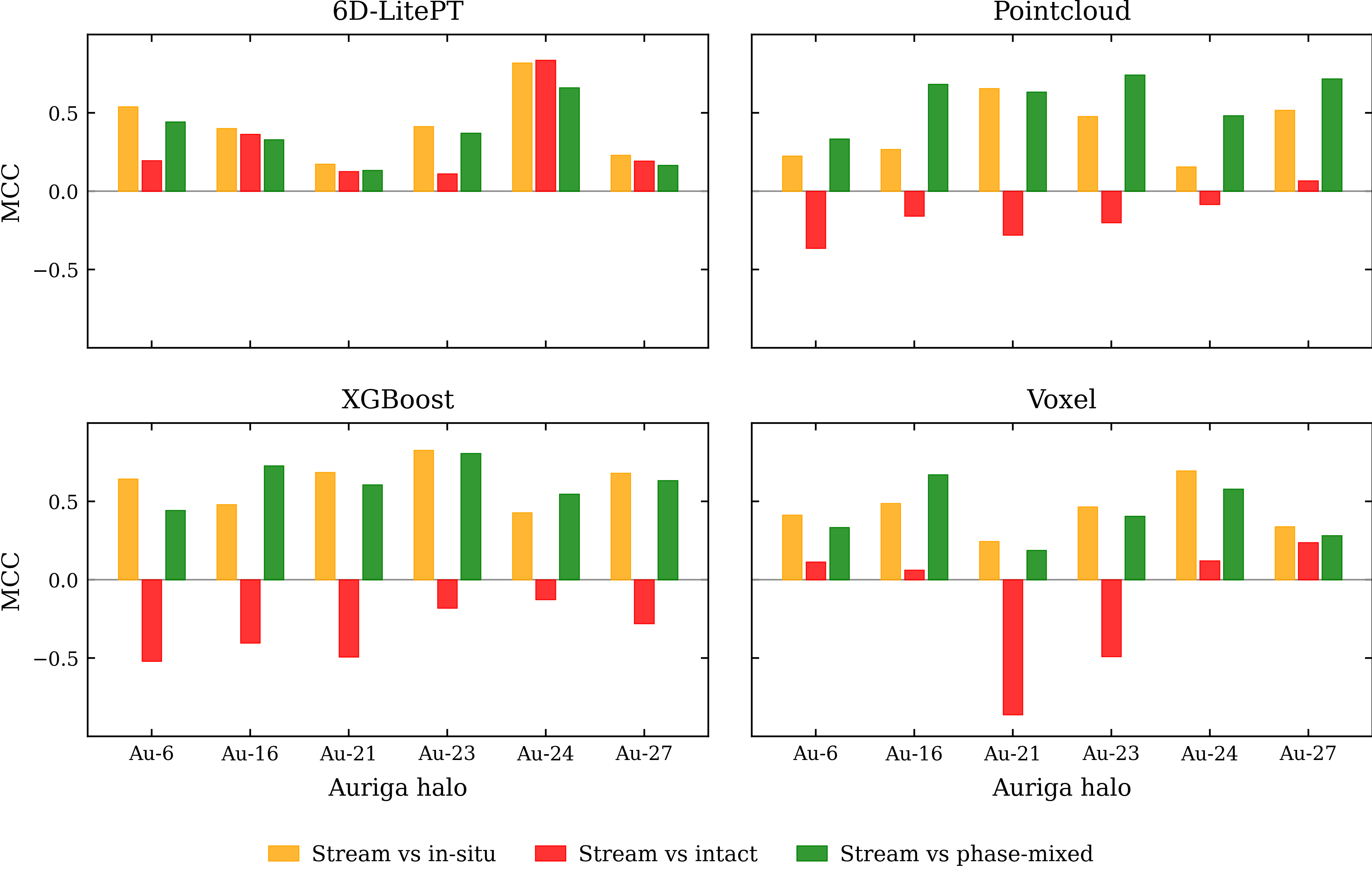}
    \caption{Category specific MCC for the four models against the six tested Auriga halos. Each comparison includes only reference stream particles (positive class) and particles from one background category (negative class): in-situ (orange), intact (red), or phase-mixed (green). Negative MCC indicates an anticorrelation with the reference classification within the corresponding two-category population.}
    \label{fig:per_category_performance}
\end{figure*}

\begin{table}
\centering
\caption{Mean particle level classification metrics across the six Auriga level-3 halos. Metrics are calculated separately for each halo and then averaged with equal weight. The highest mean precision, recall and MCC are highlighted in bold.}
\label{tab:auriga_mean_results}

\renewcommand{\arraystretch}{1.25}
\setlength{\tabcolsep}{4pt}

\begin{tabular*}{0.95\columnwidth}{@{\extracolsep{\fill}}lccc@{}}
\toprule
Model & Precision & Recall & MCC \\
\midrule
XGBoost    & 0.253          & 0.615          & 0.306 \\
Pointcloud & 0.169          & \textbf{0.679} & 0.258 \\
Voxel      & 0.280          & 0.312          & 0.197 \\
6D-LitePT  & \textbf{0.617} & 0.256          & \textbf{0.326} \\
\bottomrule
\end{tabular*}

\end{table}

\begin{table}
\centering
\caption{Mean percentage of particles selected as streams within each reference
category across the six Auriga level-3 halos. Each percentage is calculated relative to the evaluated particle population of that category within a halo, then averaged with equal weight across halos. The stream column therefore corresponds to mean recall, while the remaining columns give category specific false positive rates.}
\label{tab:auriga_category_selection}

\begingroup
\small
\renewcommand{\arraystretch}{1.2}
\setlength{\tabcolsep}{2pt}

\begin{tabular*}{\columnwidth}{@{\extracolsep{\fill}}lrrrr@{}}
\toprule
Model & In-situ (\%) & Intact (\%) & Stream (\%) & \shortstack{Phase-\\mixed} (\%) \\
\midrule
6D-LitePT  & 0.028 & 2.044  & 25.579          & 0.581 \\
Pointcloud & 5.477 & 87.135 & \textbf{67.894} & 3.144 \\
XGBoost    & 0.716 & 93.266 & 61.534          & 1.268 \\
Voxel      & 0.461 & 47.625 & 31.197          & 1.088 \\
\bottomrule
\end{tabular*}

\endgroup
\end{table}

To assess whether the behaviour observed in Au-24 extends to other cosmological environments, we evaluate the original models across the six Auriga level-3 halos Au-6, Au-16, Au-21, Au-23, Au-24 and Au-27. Fig. \ref{fig:confusion_matrices} presents the confusion matrices, while Tables \ref{tab:auriga_mean_results} and \ref{tab:auriga_category_selection} summarise the mean classification metrics and category specific selection rates. The tabulated quantities are calculated independently for each halo and averaged with equal weight.

The primary characteristics seen in both the synthetic tests and Au-24 are once again apparent for 6D-LitePT and Pointcloud. The 6D-LitePT model achieves the highest mean precision $0.617$ and MCC $0.326$, whereas Pointcloud achieves the highest mean recall $0.679$ at lower precision $0.169$. XGBoost provides a mean recall of $0.615$, precision of $0.253$, while Voxel recovers $0.312$ of stream particles on average with a precision of $0.280$. The confusion matrices provide a complementary view of the false positive and false negative behaviour underlying these results. In comparison to the halo + structures synthetic data set, we see that the XGBoost model has a nearly identical MCC performance with $0.311$ against the synthetic data and $0.306$ against the Auriga data. However, looking at precision and recall, it has great stream finding ability in the synthetic data while having better contamination removal in the Auriga data. Pointcloud performs consistently, showing lower precision and the highest recall in both the synthetic and Auriga datasets. Voxel similarly performs consistently, with roughly equal precision and recall in both datasets, albeit at lower values in the case of Auriga. Finally, 6D-LitePT, the strongest performer against the synthetic data has the greatest decrease in performance compared to the four models. However it is once again the best performer with the strongest precision.

These results demonstrate that models trained on synthetic data can identify stream particles in an independent cosmological dataset, although performance varies across models and halos. Several factors may contribute to the decrease in performance relative to the synthetic tests. Model wise, the grid production of Voxel and 6D-LitePT models may result in sharp boundary effects. This is best seen in the 6D-LitePT filtering of Au-16 in Appendix \ref{appendix}. We see the pink and orange streams in the centre form box like edges. This effect may also lead to missed voxels where there is not enough context. Looking at the datasets next, the synthetic dataset was assembled from streams and background populations that occupied similar kinematic regimes but evolved independently, whereas the Auriga populations developed within a shared cosmological environment. Class imbalance also differs substantially. Au-6 contains only $\sim40,000$ stream particles in a sample of over $16$ million, compared with the approximately $1:10$ ratio used in the synthetic benchmark. At such low stream abundance, even a small background false positive rate can substantially reduce precision. 

A further difficulty is the distinction between the reference classes. Intact satellites undergoing early disruption may share spatial and kinematic properties with stream cores, even when their tidal features are weak or absent. Without temporal context, these similarities may make the populations difficult to distinguish. More dispersed stream populations may also be harder to distinguish from the surrounding background. Differences in progenitor properties, disruption histories and local particle sampling may also contribute. The synthetic training set spans a controlled family of structures, whereas Auriga presents a wider range of cosmological environments. Moreover, fixed neighbour counts and voxel scales can sample different physical extents or mixtures of populations when particle densities differ. Addressing these issues may benefit from training on more representative populations. However, tailoring the training too closely to a particular simulation or population could limit generalisation to other environments. 

Table \ref{tab:auriga_category_selection} shows that confusion with intact satellites extends beyond the Au-24 example. Pointcloud and XGBoost select, on average, $87.135\%$ and $93.266\%$ of intact-system particles, respectively, exceeding their corresponding stream recovery rates. Both models nevertheless reject most in-situ and phase-mixed particles, indicating substantially greater difficulty distinguishing streams from intact satellites than from these background populations. In contrast, 6D-LitePT selects only $2.044\%$ of intact particles, alongside $0.028\%$ of in-situ and $0.581\%$ of phase-mixed particles, consistent with its greater precision but more limited completeness. Voxel exhibits similar intact contamination with $47.625\%$ while being the second best filter of in-situ and phase-mixed structures behind 6D-LitePT.

These results reinforce the contrasting selection behaviours illustrated by Au-24: Pointcloud favours completeness but struggles to reject intact satellites, while 6D-LitePT provides a cleaner selection at the cost of missing a larger fraction of the stream population. The low category specific selection rates for in-situ and phase-mixed particles achieved by the models demonstrate useful background rejection, although the large sizes of these populations mean that even small false positive rates can contribute appreciable contamination. Distinguishing streams from intact systems remains a major limitation for Pointcloud and XGBoost, while improving recovery without sacrificing background rejection is the corresponding challenge for 6D-LitePT. These complementary strengths favour 6D-LitePT when limiting contamination is the priority and Pointcloud when completeness is prioritised alongside subsequent candidate verification.

Figure \ref{fig:per_category_performance} quantifies the ability of each model to distinguish streams from background particles for each reference category. For the stream versus intact comparison, MCC is negative in all six halos for XGBoost, in five for Pointcloud, and in two for Voxel. This anticorrelation indicates that, in these cases, the model selects a larger fraction of intact particles than of reference stream particles. 6D-LitePT maintains positive MCC against intact systems in all six halos, though performing the weakest when compared with the other categories. In particular, Au-6 and Au-23 show reduced performance similar to the other models, albeit while still maintaining a positive MCC. All four models attain positive MCC against both the in-situ and phase-mixed populations in every halo, with scores generally higher than in the intact comparison. One possible explanation is that compact, kinematically cold regions of intact satellites can resemble locally coherent parts of a stream core. This interpretation is consistent with the overlap in velocity dispersion seen for some intact and stream systems in Au-24 (Fig. \ref{fig:auriga_system_properties}). Phase mixing disperses debris in phase space, while the in-situ population is not dominated by the coherent debris of individual accreted progenitors. These differences may help explain why all four models achieve positive MCC against both populations across the six halos.

The transferability of the model is displayed in its performance against the Auriga simulation suite. As discussed previously, several factors must be considered when evaluating performance. The distributions of progenitor properties, disruption histories, particle densities and spatial and kinematic scales, together with the different reference classifications must all be considered. These considerations also apply to observational catalogues, where survey selection, extinction, foreground contamination, measurement uncertainties and incomplete phase space information further alter the sampled populations and their apparent structure. As a result, with the hindsight of the models' performances here, the application to observational data would require alterations to account for these additional factors for a stronger classification. Where necessary, retraining or adaptation of the input representation would be needed, followed by independent validation using known streams and injected mock structures to produce a more robust model against a more difficult dataset.

\section{Conclusion}
\label{sec:conclusion}

In this work, we constructed an ML based training, validation and testing data pipeline for the identification of stellar streams in Milky Way-like galactic halos. We compare four separate frameworks: 1. Enriched feature decision-tree operated with gradient boosting algorithm XGBoost and deep learning algorithms, 2. Convolutional Neural Network (CNN) based Voxel segmentation, 3. PointNet++ inspired Pointcloud segmentation, 4. Transformer based hybrid 6D-LitePT model. We test the models against class imbalanced scenarios, firstly with only a Milky Way-like host halo, and second, with the addition of extra non-stream-like coherent structures. We found that the Pointcloud model excels in the halo only scenario but experiences a drastic decrease in performance with the inclusion of additional structures. LitePT remained the most consistent, providing excellent performance in both scenarios. XGBoost performed similarly to Pointcloud with a drop off in performance. The Voxel model performed consistently with similar precision and recall.

We then applied the original synthetic trained models to six independent Auriga level-3 halos, using published system classifications as particle level reference labels. This evaluation provides a quantitative assessment of transfer to cosmological simulations. Across the suite, 6D-LitePT achieved the highest mean precision ($0.617$) and MCC ($0.326$), while Pointcloud achieved the highest mean recall ($0.679$). These results identify complementary strengths: 6D-LitePT provides a cleaner selection, whereas Pointcloud recovers a larger fraction of the reference stream population with substantially greater contamination. XGBoost maintained a consistent MCC between the halo + structures synthetic and Auriga datasets. Voxel performed the overall poorest against Auriga data, despite having the second strongest precision.

The transfer results also reflect differences between the synthetic and cosmological populations. The fixed class balance of the synthetic benchmark differs from the varying stream abundance across Auriga, affecting precision even at unchanged recall and background false positive rates. Differences in progenitor properties, disruption histories, particle sampling and phase space distributions can additionally affect the learned classifications. The results therefore establish the measured capabilities and limitations of the adopted pipelines under these changes. Future development should target the limitations identified for each pipeline, including discrimination between streams and intact satellites, improved recovery of diffuse debris, and rejection of the host background. Representative training populations and explicit treatment of the intended classification target provide avenues for investigating these improvements. Application to observations would additionally require accounting for survey selection, foreground contamination, measurement uncertainties and incomplete phase space information through realistic mock stellar catalogues and independent validation. Extending the models to instance segmentation, so that individual streams are distinguished from one another, is a further direction. The present framework provides a quantitative basis for these developments and demonstrates the potential of identifying stellar stream structures from exclusively 6D phase space information.

\section*{Acknowledgements}
We are grateful for the feedback and comments provided by the Gravitational Astrophysics group in the University of Sydney's School of Physics. We would also like to thank the anonymous reviewer for their feedback on the paper. This research was undertaken with the assistance of resources from the National Computational Infrastructure (NCI Australia), an NCRIS enabled capability supported by the Australian Government and the Sydney Informatics Hub, a Core Research Facility of the University of Sydney. VE is supported by the University of Sydney Postgraduate Award. SGB is supported by Australian Academy of Technological Sciences and Engineering's Elevate scholarship. AI was used in the production of the ML as well as editing and figure codes used in this paper. We have used simulations from the Auriga Project public data release \cite{Grand_2024} available at \url{https://wwwmpa.mpa-garching.mpg.de/auriga/data.html}.

%%%%%%%%%%%%%%%%%%%%%%%%%%%%%%%%%%%%%%%%%%%%%%%%%%
\section*{Data Availability}

The code can be found at \url{https://github.com/Viraj-E/ML-stellar-stream-classification}. The synthetic simulation data utilised in the testing pipeline in the paper will be shared upon reasonable request to the corresponding author.



\bibliographystyle{mnras}
\bibliography{main} % if your bibtex file is called example.bib

%\begin{thebibliography}{99}
%\end{thebibliography}

%%%%%%%%%%%%%%%%%%%%%%%%%%%%%%%%%%%%%%%%%%%%%%%%%%

%%%%%%%%%%%%%%%%% APPENDICES %%%%%%%%%%%%%%%%%%%%%

\appendix

\section{Additional Auriga halo results}
\label{appendix}
This appendix presents the remaining five level-3 Auriga halos, complementing the Au-24 comparison in the main text. For each halo, we show the catalogue streams alongside the particles selected by the four models in the $y-z$ plane and report particle level performance metrics and selection rates by reference category.

\begin{figure*}
    \centering
    \includegraphics[width=\linewidth]{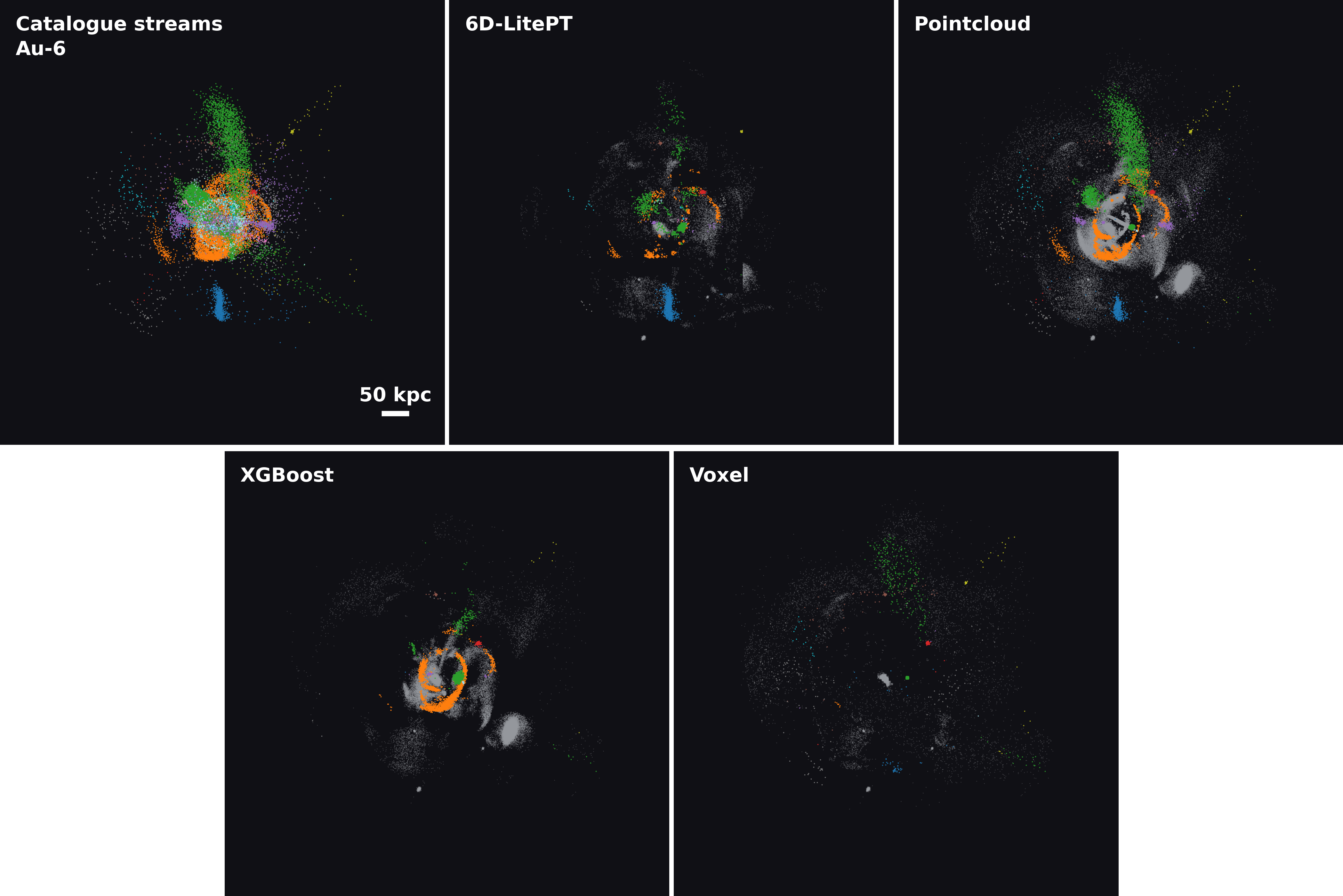}
    \caption{Comparison of the catalogue stream population and the particles selected by the original synthetic trained models in Au-6 at level-3 resolution at $z=0$, projected onto the $y$--$z$ plane. Coloured particles belong to the reference stream systems, using the same colours as the first panel, while grey particles belong to non-stream reference populations.}
    \label{fig:auriga6_stream_comparison}

\begin{minipage}{\textwidth}
\centering
\captionof{table}{Performance and category specific selection rates for Au-6 at level-3 resolution.}
\label{tab:auriga6_combined}

\renewcommand{\arraystretch}{1.3}
\setlength{\tabcolsep}{6pt}
\begin{tabular*}{\textwidth}{@{\extracolsep{\fill}}lccccccc@{}}
\hline
\multirow{2}{*}{Model}
& \multicolumn{3}{c}{Classification metrics}
& \multicolumn{4}{c}{Selected within reference category} \\
\cline{2-8}
& Precision & Recall & MCC
& In-situ (\%) & Intact (\%) & Stream (\%) & Phase-mixed (\%) \\
\hline
XGBoost
& 0.022 & 0.466 & 0.091
& 0.018 & 97.918 & 46.578 & 1.294 \\
Pointcloud
& 0.019 & \textbf{0.494} & 0.087
& 1.253 & 94.733 & \textbf{49.406} & 3.192 \\
Voxel
& 0.124 & 0.192 & 0.151
& 0.007 & 5.789 & 19.210 & 0.271 \\
6D-LitePT
& \textbf{0.161} & 0.330 & \textbf{0.228}
& 0.013 & 7.030 & 32.969 & 0.442 \\
\hline
\end{tabular*}
\end{minipage}
\end{figure*}

\begin{figure*}
    \centering
    \includegraphics[width=\linewidth]{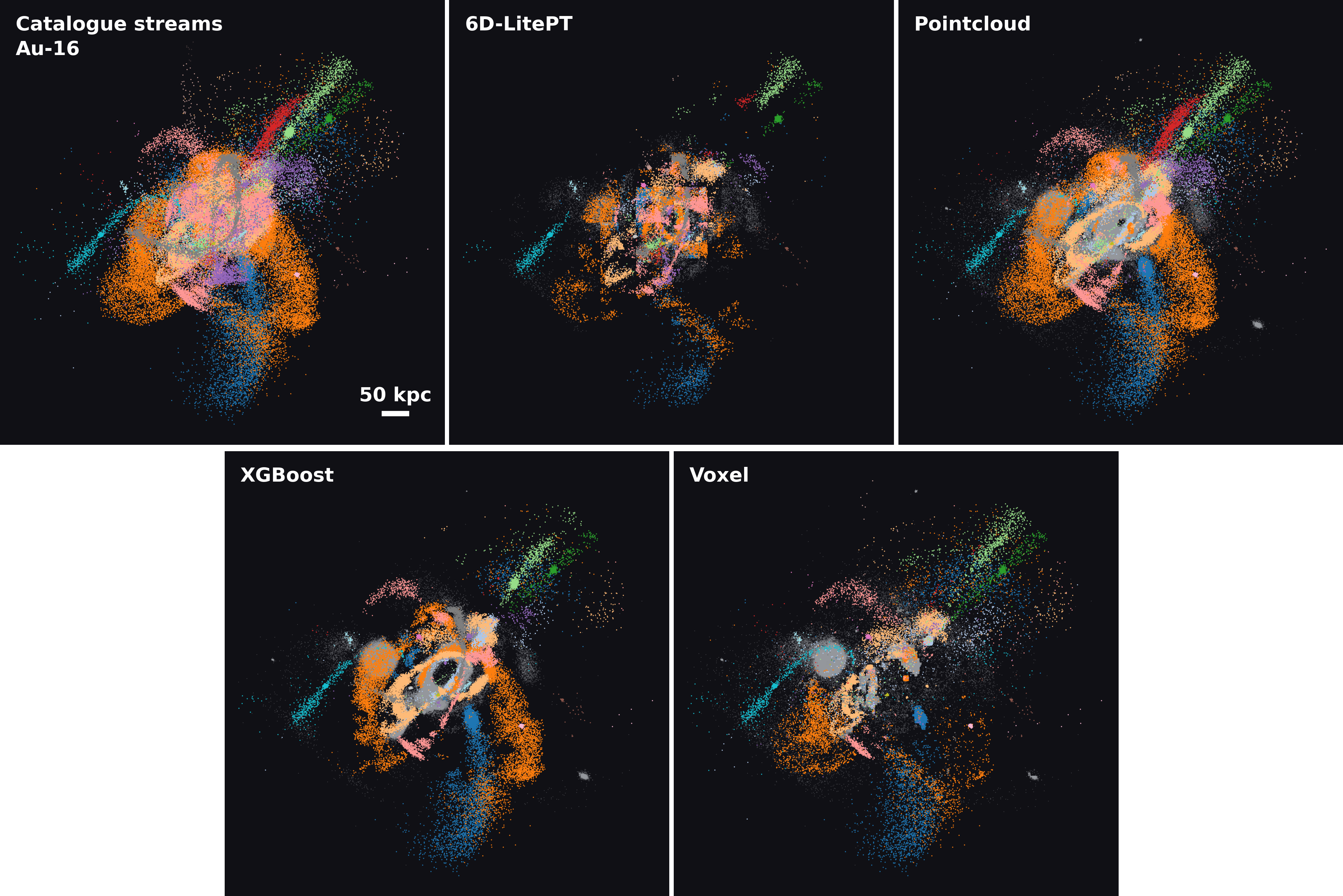}
    \caption{Comparison of the catalogue stream population and the particles selected by the original synthetic trained models in Au-16 at level-3 resolution at $z=0$, projected onto the $y$--$z$ plane. Coloured particles belong to the reference stream systems, using the same colours as the first panel, while grey particles belong to non-stream reference populations.}
    \label{fig:auriga16_stream_comparison}

\begin{minipage}{\textwidth}
\centering
\captionof{table}{Performance and category specific selection rates for Au-16 at level-3 resolution.}
\label{tab:auriga16_combined}

\renewcommand{\arraystretch}{1.3}
\setlength{\tabcolsep}{6pt}
\begin{tabular*}{\textwidth}{@{\extracolsep{\fill}}lccccccc@{}}
\hline
\multirow{2}{*}{Model}
& \multicolumn{3}{c}{Classification metrics}
& \multicolumn{4}{c}{Selected within reference category} \\
\cline{2-8}
& Precision & Recall & MCC
& In-situ (\%) & Intact (\%) & Stream (\%) & Phase-mixed (\%) \\
\hline
XGBoost
& 0.094 & 0.725 & 0.229
& 3.220 & 98.324 & 72.459 & 3.078 \\
Pointcloud
& 0.063 & \textbf{0.838} & 0.189
& 15.639 & 94.843 & \textbf{83.814} & 8.422 \\
Voxel
& 0.128 & 0.678 & 0.268
& 2.611 & 59.437 & 67.826 & 3.913 \\
6D-LitePT
& \textbf{0.694} & 0.175 & \textbf{0.344}
& 0.026 & 0.330 & 17.515 & 0.893 \\
\hline
\end{tabular*}
\end{minipage}
\end{figure*}

Au-6 (Fig. \ref{fig:auriga6_stream_comparison}) has the lowest MCC averaged across the four models. 6D-LitePT achieves the highest MCC for this halo, while Pointcloud recovers the largest fraction of catalogue stream particles. In Au-16 (Fig. \ref{fig:auriga16_stream_comparison}), the same trade off is more pronounced. Pointcloud recovers approximately $84\%$ of stream particles but also selects most intact particles, while 6D-LitePT achieves the highest precision and MCC. Its recovered streams show localised gaps, as discussed in Section \ref{subsec:auriga_suite_inference}.

\begin{figure*}
    \centering
    \includegraphics[width=\linewidth]{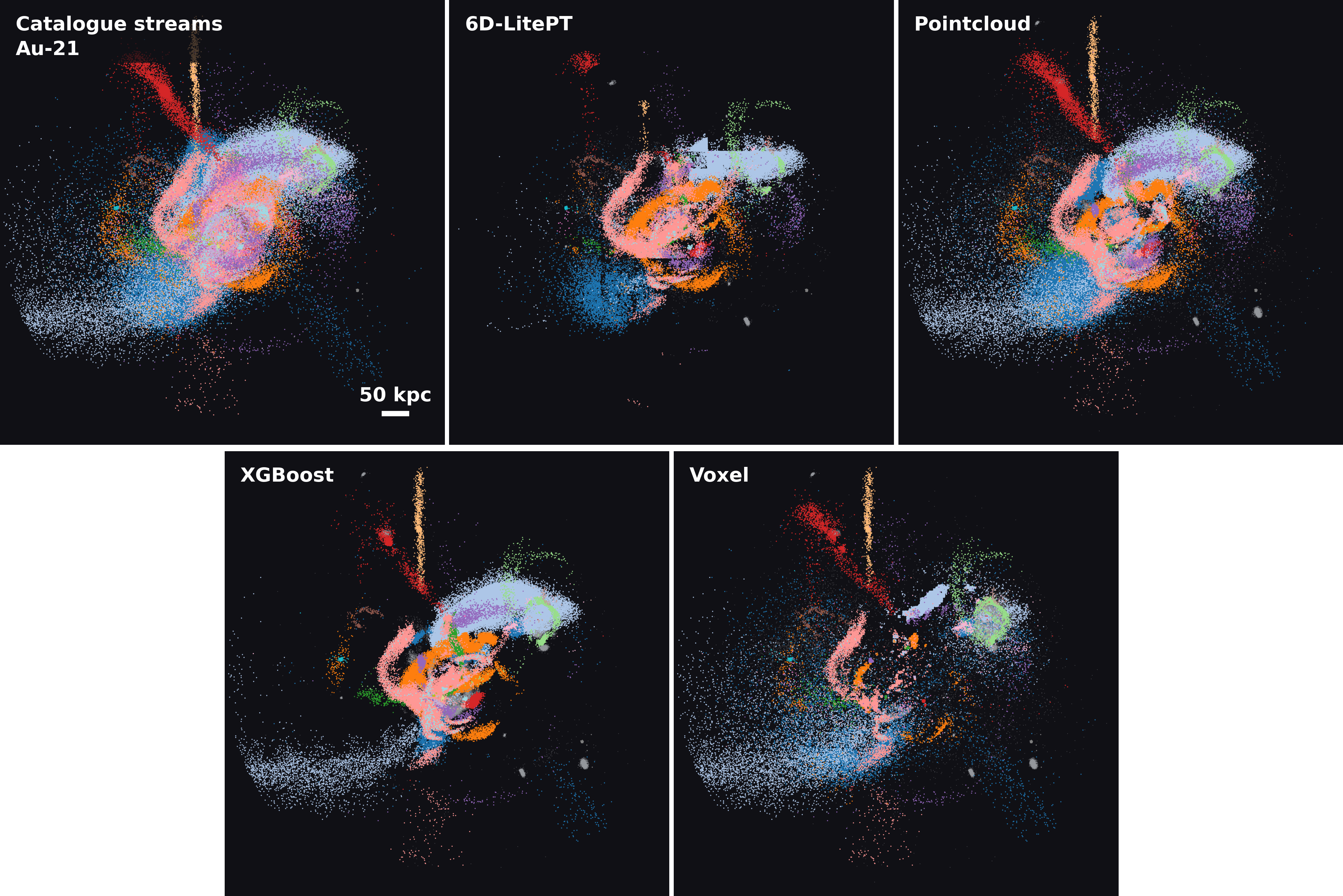}
    \caption{Comparison of the catalogue stream population and the particles selected by the original synthetic trained models in Au-21 at level-3 resolution at $z=0$, projected onto the $y$--$z$ plane. Coloured particles belong to the reference stream systems, using the same colours as the first panel, while grey particles belong to non-stream reference populations.}
    \label{fig:auriga21_stream_comparison}

\begin{minipage}{\textwidth}
\centering
\captionof{table}{Performance and category specific selection rates for Au-21 at level-3 resolution.}
\label{tab:auriga21_combined}

\renewcommand{\arraystretch}{1.3}
\setlength{\tabcolsep}{6pt}
\begin{tabular*}{\textwidth}{@{\extracolsep{\fill}}lccccccc@{}}
\hline
\multirow{2}{*}{Model}
& \multicolumn{3}{c}{Classification metrics}
& \multicolumn{4}{c}{Selected within reference category} \\
\cline{2-8}
& Precision & Recall & MCC
& In-situ (\%) & Intact (\%) & Stream (\%) & Phase-mixed (\%) \\
\hline
XGBoost
& 0.280 & 0.526 & 0.275
& 0.221 & 94.638 & 52.616 & 0.149 \\
Pointcloud
& 0.311 & \textbf{0.569} & \textbf{0.320}
& 1.751 & 82.298 & \textbf{56.899} & 0.452 \\
Voxel
& 0.052 & 0.073 & -0.082
& 0.042 & 93.649 & 7.337 & 0.287 \\
6D-LitePT
& \textbf{0.817} & 0.038 & 0.161
& 0.030 & 0.375 & 3.771 & 0.134 \\
\hline
\end{tabular*}
\end{minipage}
\end{figure*}

\begin{figure*}
    \centering
    \includegraphics[width=\linewidth]{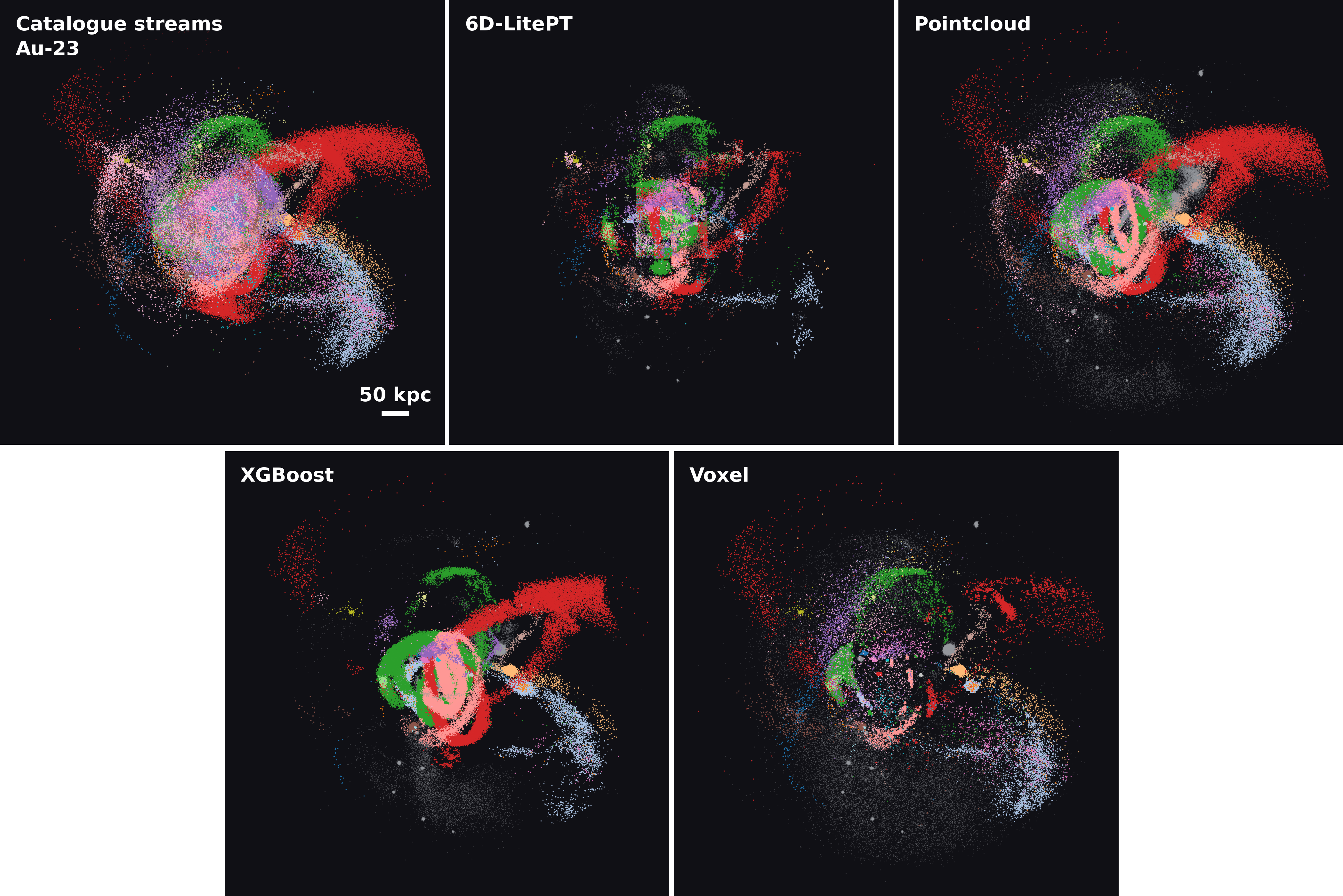}
    \caption{Comparison of the catalogue stream population and the particles selected by the original synthetic trained models in Au-23 at level-3 resolution at $z=0$, projected onto the $y$--$z$ plane. Coloured particles belong to the reference stream systems, using the same colours as the first panel, while grey particles belong to non-stream reference populations.}
    \label{fig:auriga23_stream_comparison}

\begin{minipage}{\textwidth}
\centering
\captionof{table}{Performance and category specific selection rates for Au-23 at level-3 resolution.}
\label{tab:auriga23_combined}

\renewcommand{\arraystretch}{1.3}
\setlength{\tabcolsep}{6pt}
\begin{tabular*}{\textwidth}{@{\extracolsep{\fill}}lccccccc@{}}
\hline
\multirow{2}{*}{Model}
& \multicolumn{3}{c}{Classification metrics}
& \multicolumn{4}{c}{Selected within reference category} \\
\cline{2-8}
& Precision & Recall & MCC
& In-situ (\%) & Intact (\%) & Stream (\%) & Phase-mixed (\%) \\
\hline
XGBoost
& 0.790 & \textbf{0.732} & \textbf{0.755}
& 0.131 & 99.006 & \textbf{73.217} & 0.741 \\
Pointcloud
& 0.317 & 0.706 & 0.454
& 3.824 & 99.972 & 70.560 & 2.059 \\
Voxel
& 0.586 & 0.235 & 0.362
& 0.037 & 99.748 & 23.522 & 0.782 \\
6D-LitePT
& \textbf{0.839} & 0.184 & 0.387
& 0.028 & 4.357 & 18.409 & 0.374 \\
\hline
\end{tabular*}
\end{minipage}
\end{figure*}

\begin{figure*}
    \centering
    \includegraphics[width=\linewidth]{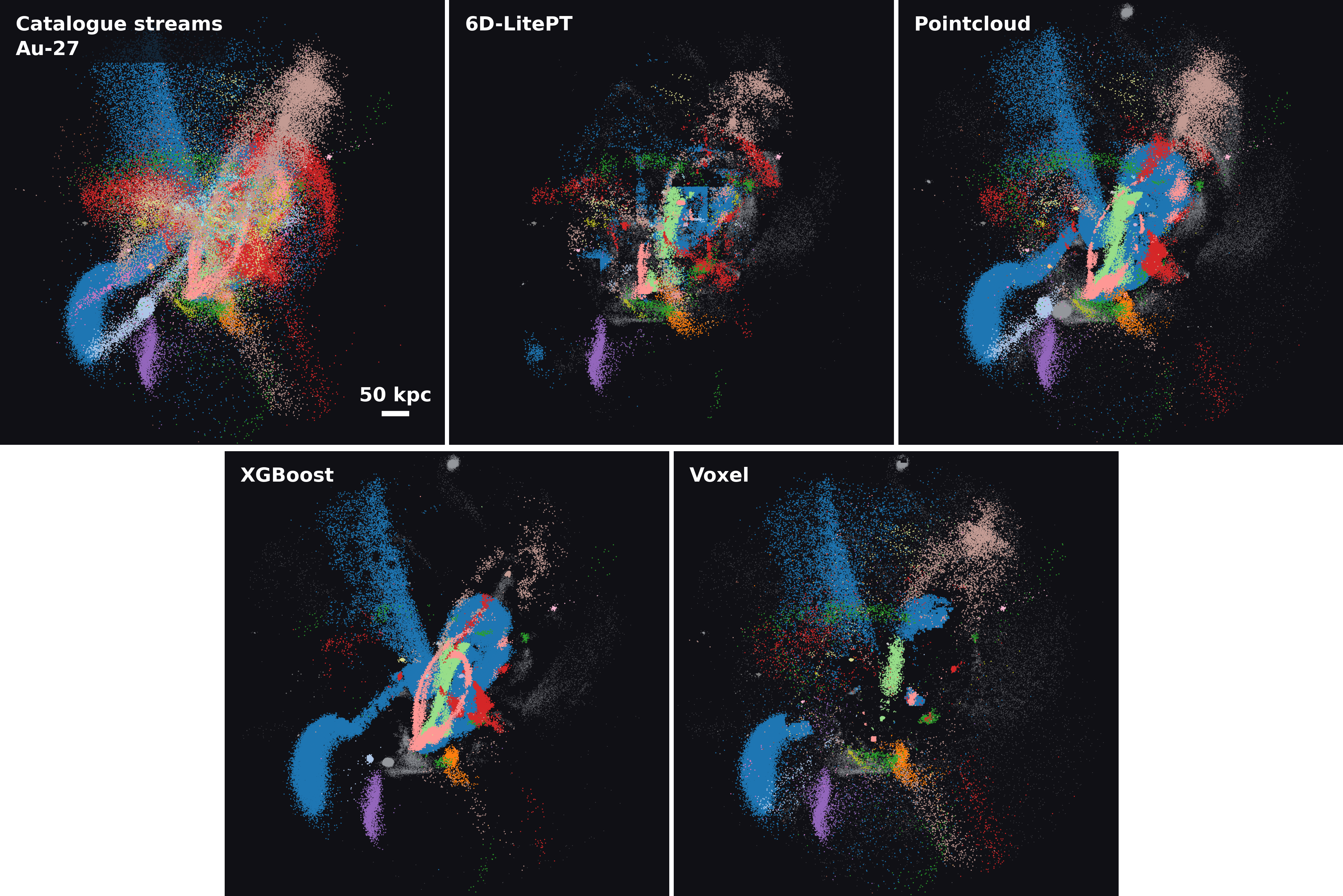}
    \caption{Comparison of the catalogue stream population and the particles selected by the original synthetic trained models in Au-27 at level-3 resolution at $z=0$, projected onto the $y$--$z$ plane. Coloured particles belong to the reference stream systems, using the same colours as the first panel, while grey particles belong to non-stream reference populations.}
    \label{fig:auriga27_stream_comparison}

\begin{minipage}{\textwidth}
\centering
\captionof{table}{Performance and category specific selection rates for Au-27 at level-3 resolution.}
\label{tab:auriga27_combined}

\renewcommand{\arraystretch}{1.3}
\setlength{\tabcolsep}{6pt}
\begin{tabular*}{\textwidth}{@{\extracolsep{\fill}}lccccccc@{}}
\hline
\multirow{2}{*}{Model}
& \multicolumn{3}{c}{Classification metrics}
& \multicolumn{4}{c}{Selected within reference category} \\
\cline{2-8}
& Precision & Recall & MCC
& In-situ (\%) & Intact (\%) & Stream (\%) & Phase-mixed (\%) \\
\hline
XGBoost
& 0.312 & 0.489 & 0.361
& 0.061 & 75.976 & 48.896 & 0.485 \\
Pointcloud
& 0.291 & \textbf{0.629} & \textbf{0.396}
& 3.429 & 56.167 & \textbf{62.933} & 1.243 \\
Voxel
& \textbf{0.729} & 0.130 & 0.298
& 0.042 & 1.331 & 12.979 & 0.656 \\
6D-LitePT
& 0.643 & 0.065 & 0.196
& 0.047 & 0.055 & 6.461 & 0.788 \\
\hline
\end{tabular*}
\end{minipage}
\end{figure*}

In Au-21 (Fig. \ref{fig:auriga21_stream_comparison}), Pointcloud achieves the highest MCC, while the Voxel model obtains a negative MCC and selects a much larger fraction of intact particles than stream particles. Au-23 (Fig. \ref{fig:auriga23_stream_comparison}) has the highest MCC averaged across the four models, with XGBoost achieving the highest individual MCC of $0.755$ across all models and Auriga halos. Finally, in Au-27 (Fig. \ref{fig:auriga27_stream_comparison}), Pointcloud achieves the highest recall and MCC, while the Voxel model achieves the highest precision but recovers only a small fraction of the catalogue stream particles.

%%%%%%%%%%%%%%%%%%%%%%%%%%%%%%%%%%%%%%%%%%%%%%%%%%

% Don't change these lines
\bsp	% typesetting comment
\label{lastpage}
\end{document}